%% file: EGM-24-002_temp.tex
\pdfoutput=1
\documentclass[11pt,twoside,a4paper,cmspaper,final,collab]{cms-tdr}
\def\svnVersion{e6178ef}\def\svnDate{2026/09/08}\def\cmsCernNoTag{CERN-EP-2026-092}\def\cmsCernDate{\today}\def\cmsMessage{Published in Physical Review D as \href{http://dx.doi.org/10.1103/hb72-vjj6}{\doi{10.1103/hb72-vjj6}.}}
\begin{document}\cmsNoteHeader{EGM-24-002}

\ifthenelse{\boolean{cms@external}}{\providecommand{\cmsLeft}{upper\xspace}}{\providecommand{\cmsLeft}{left\xspace}}
\ifthenelse{\boolean{cms@external}}{\providecommand{\cmsRight}{lower\xspace}}{\providecommand{\cmsRight}{right\xspace}}

\newcommand{\gammaL}{\ensuremath{\gamma_{\mathrm{L}}}\xspace}
\newcommand{\PeME}{\ensuremath{\Pe_{\mathrm{ME}}}\xspace}
\newcommand{\PY}{{\HepParticle{Y}{}{}}\Xspace}
\newcommand{\mass}{\ensuremath{m}\xspace}
\newcommand{\varMass}[1]{\ensuremath{{\mass}_{#1}}\xspace}
\newcommand{\Mee}{\ensuremath{\varMass{\Pe\Pe}}\xspace}
\newcommand{\MX}{\ensuremath{\varMass{\PX}}\xspace}
\newcommand{\MY}{\ensuremath{\varMass{\PY}}\xspace}
\newcommand{\Mmmg}{\ensuremath{\varMass{\PGm\PGm\PGg}}\xspace}
\newcommand{\Mmm}{\ensuremath{\varMass{\PGm\PGm}}\xspace}
\newcommand{\Xzero}{\ensuremath{X_{0}}\xspace}
\newcommand{\dEtaInSeed}{\ensuremath{\Delta\eta_{\mathrm{in}}^{\text{seed}}}\xspace}
\newcommand{\dUInfxf}{\ensuremath{\Delta u}\xspace}
\newcommand{\dVInfxf}{\ensuremath{\Delta v}\xspace}
\newcommand{\covIxIx}[3]{\ensuremath{\mathrm{cov}^{#3}_{i#1 i#2}}\xspace}
\newcommand{\covfxfIxIx}[2]{\ensuremath{\covIxIx{#1}{#2}{5\times5}}\xspace}
\newcommand{\covfxfIeIe}{\ensuremath{\covfxfIxIx{\eta}{\eta}}\xspace}
\newcommand{\alphaTrack}{\ensuremath{\alpha_{\text{track}}}\xspace}
\newcommand{\alphafxf}{\ensuremath{\alpha_{5\times5}}\xspace}
\newcommand{\ufxf}{\ensuremath{\mathrm{U}_{5\times5}}\xspace}
\newcommand{\varI}[1]{\ensuremath{{#1}_{i}}\xspace}
\newcommand{\varCoGufxf}[1]{\ensuremath{\overline{#1}_{\mathrm{U_{5\times5}}}}\xspace}
\newcommand{\Eufxf}{\ensuremath{E_{\mathrm{U}_{5\times5}}}\xspace}
\newcommand{\dXInUfxf}[1]{\ensuremath{\Delta#1_{\mathrm{in}}^{5\times5}}\xspace}
\newcommand{\dXIn}[1]{\ensuremath{\Delta#1_{\mathrm{in}}}\xspace}
\newcommand{\HoverE}{\ensuremath{H/E}\xspace}
\newcommand{\EoverP}{\ensuremath{E/p}\xspace}
\newcommand{\ETfxf}{\ensuremath{E_{\mathrm{T}}^{5\times5}}\xspace}
\newcommand{\dxy}{\ensuremath{d_{xy}}\xspace}
\newcommand{\Lxy}{\ensuremath{L_{xy}}\xspace}
\newcommand{\dzero}{\ensuremath{d_{0}}\xspace}
\newcommand{\CItwoSig}{\ensuremath{\mathrm{CI}_{0.95}}\xspace}
\newcommand{\alphaTwoD}{\ensuremath{\alpha_{\mathrm{2D}}}\xspace}

\newlength\cmsTabSkip\setlength{\cmsTabSkip}{1.3ex}

\newlength\cmsFigWidth
\ifthenelse{\boolean{cms@external}}{
  \setlength\cmsFigWidth{0.49\textwidth}}{\setlength\cmsFigWidth{0.6\textwidth}}

\cmsNoteHeader{EGM-24-002}
\title{Highly boosted dielectron identification in proton-proton collisions at \texorpdfstring{$\sqrt{s} = 13\TeV$}{sqrt(s) = 13 TeV}}

\date{\today}

\abstract{
  A new technique is developed to identify dielectrons (\EE) with Lorentz boost $\gammaL > 20$ that produce one single merged cluster in the electromagnetic calorimeter of the CMS detector. The identification uses two multivariate models: one for the case where both electron tracks are reconstructed, and another where only one of the tracks is reconstructed. The efficiency is determined using proton-proton collision data collected at a center-of-mass energy of 13\TeV. Boosted \PJGy mesons decaying into \EE pairs are used to estimate the efficiency of the model with two tracks, yielding an overall efficiency of 80\%. The $\PZ\to\MM\PGg$ events, where the photon converts into a collimated dielectron, are used for the model with a single track, yielding an efficiency of about 60\%. A dedicated energy correction for dielectron candidates is also developed using $\PBpm\to\PJGy\PKpm\to\EE\PKpm$ data.
}

\hypersetup{
pdfauthor={CMS Collaboration},
pdftitle={Highly boosted dielectron identification in proton-proton collisions at sqrt(s) = 13 TeV},
pdfsubject={CMS},
pdfkeywords={merged electron, high-energy, boosted signature, B Parking}}

\maketitle

\section{Introduction}

The electron and photon reconstruction and identification (ID) techniques developed by the CMS Collaboration at the CERN LHC provide excellent energy resolution and efficiency, and their performance was rigorously demonstrated using LHC Run 2 proton-proton collision data at $\sqrt{s} = 13\TeV$ collected during 2016--2018~\cite{CMS:2020uim}, corresponding to an integrated luminosity of 138\fbinv. The techniques are designed to deliver the best possible performance for the well-isolated prompt electrons and photons typically used in precision measurements of standard model (SM) processes and in many searches for physics beyond the SM (BSM).

However, numerous BSM models predict light bosons associated with new heavy resonances, which may subsequently decay into a dielectron (\EE)~\cite{ZprimePheno,2HDMreview,darkphoton}, whose angular separation can be small.
This can make the individual identification of each electron difficult when the opening angle is similar or smaller than the calorimeter granularity or the size of its effective Moli{\`e}re radius. 
Also, in order to obtain the best possible energy resolution, electron reconstruction algorithms are designed to identify associated bremsstrahlung photons near electron candidates and include them in the electron energy. This can result in the merging of two nearby electrons.

In this paper, we present an algorithm for Run 2 data to identify merged electrons (\PeME) with an arbitrary dielectron invariant mass, \Mee.
The CMS Collaboration has previously developed algorithms to identify merged photon clusters~\cite{PhysRevD.108.052002,EXO-22-022} using images of the energy deposits recorded in the calorimeter. The ATLAS Collaboration has previously developed an algorithm to identify \PeME from the decay of highly boosted on-shell \PZ bosons with $60 < \Mee < 120\GeV$~\cite{ATLASZee}.

The algorithm has two types of \PeME, with (``two-track'') and without (``single-track'') a secondary associated track. Each type is identified with its own multivariate model. The two-track model relies on the geometrical compatibility between the merged cluster and the two electron tracks. Its efficiency is validated using boosted \PJGy mesons decaying to \EE since they provide an abundant number of \EE pairs with Lorentz boost $\gammaL > 20$, in contrast to the \PZ boson that requires about 2\TeV of energy to reach the same level of \gammaL. The single-track model mainly utilizes the ratio between the merged cluster's energy and the electron track's momentum. Its efficiency is measured using $\PZ\to\MM\PGg$ events with a photon conversion, because the \EE pair from a converted photon often has only a single track reconstructed due to the insufficient separation, mimicking the feature of the single-track \PeME. For the two-track model, a dedicated energy correction is also developed to match the energy description of the merged cluster in simulation to that in data. As the cluster from $\PJGy\to\EE$ is merged, $\PBpm\to\PJGy\PKpm\to\EE\PKpm$ decays are used to construct a resonance structure to derive the energy scale and resolution of the merged cluster. The method described in this paper is used to identify \PeME produced in the decay of a hypothetical light BSM boson in data collected by the CMS detector~\cite{EXO24006}. The new \PeME ID significantly improves the sensitivity in the electron channel within the phase space with Lorentz boost $\gammaL > 20$ by reducing backgrounds originating from isolated single electrons.

\section{The CMS detector}

The CMS apparatus~\cite{CMS:2008xjf,CMS:2023gfb} is a multipurpose, nearly hermetic detector, designed to trigger on~\cite{CMS:2020cmk,CMS:2016ngn,CMS:2024aqx} and identify electrons, muons, photons, and (charged and neutral) hadrons~\cite{CMS:2020uim,CMS:2018rym,CMS:2014pgm}. Its central feature is a superconducting solenoid of 6\unit{m} internal diameter, providing a magnetic field of 3.8\unit{T}. Within the solenoid volume are a silicon pixel and strip tracker, a lead tungstate crystal electromagnetic calorimeter (ECAL), and a brass and scintillator hadron calorimeter (HCAL), each composed of a barrel and two endcap sections.

The silicon tracker used in 2016 measured charged particles within the pseudorapidity range $\abs{\eta} < 2.5$. The pixel tracker had three (four) barrel layers and two (three) disks on each endcap region in 2016 (2017 and 2018)~\cite{Phase1Pixel}. The strip tracker has ten layers in the barrel region with four layers in the tracker inner barrel and six layers in the tracker outer barrel. The inner barrel is supplemented with three tracker inner disks at each end. The strip tracker endcap consists of nine disks that cover up to $\abs{\eta} < 2.5$.

The ECAL consists of 75\,848 lead tungstate crystals, which provide coverage in $\abs{\eta} < 1.48$ in a barrel region (EB) and $1.48 < \abs{\eta} < 3.0$ in two endcap regions (EE). The EB is divided by 360 in azimuthal angle $\phi$ and 170 in $\eta$. Each crystal's size corresponds to $0.0174\times0.0174$ in $\eta\times\phi$ or $22\times22\mm^{2}$ at the front face of the crystal and $26\times26\mm^{2}$ at the rear face. The Moli{\`e}re radius of lead tungstate is 22\mm. The crystal length is 230\mm, corresponding to 25.8 radiation lengths (\Xzero). The centers of the front faces are located at a 1.29\unit{m} cylindrical radius from the beam axis. Preshower detectors consisting of two planes of silicon sensors interleaved with a total of 3\Xzero of lead are located in front of each EE detector. The HCAL barrel (HB) has a symmetric 72-fold segmentation in the $\phi$ direction and 32-fold in $\eta$. Each HB tower has a lateral dimension of $0.087\times0.087$ in $\eta\times\phi$. The inner radius of the HB is approximately 1.8\unit{m}. Forward calorimeters extend the pseudorapidity coverage provided by the barrel and endcap detectors. Muons are reconstructed in gas-ionization detectors embedded in the steel flux-return yoke outside the solenoid. A more detailed description of the CMS detector, together with a definition of the coordinate system used and the relevant kinematic variables, can be found in Refs.~\cite{CMS:2008xjf,CMS:2023gfb}. 

A particle-flow (PF) algorithm~\cite{CMS:2017yfk} aims to reconstruct and identify each individual particle in an event, with an optimized combination of information from the various elements of the CMS detector. The charged-particle trajectory reconstruction is seeded by two or three consecutive hits in the pixel tracker. The trajectory finding and fitting are performed by the Kalman filter (KF) algorithm, which extrapolates the seed trajectory to the next layer along the expected trajectory. It uses the most compatible hit in the next layer to update the trajectory state iteratively~\cite{CMS:2014pgm}. The energy of photons is obtained from the ECAL measurement. The energy of electrons is determined from a combination of the electron momentum at the primary interaction vertex (PV), as determined by an alternative track reconstruction using the Gaussian sum filter (GSF) algorithm~\cite{Adam_2005}, the energy of the corresponding ECAL cluster, and the energy sum of all bremsstrahlung photons spatially compatible with originating from the electron track. The energy of muons is obtained from the curvature of the corresponding KF track. The energy of charged hadrons is determined from a combination of their KF track momenta and the matching ECAL and HCAL energy deposits, corrected for the calorimeter response to hadronic showers. Finally, the energy of neutral hadrons is obtained from the corresponding corrected ECAL and HCAL energies.

For the standard reconstruction algorithm, the electron momentum resolution for electrons with the transverse momentum $\pt\approx45\GeV$ from $\PZ\to\EE$ decays ranges from 1.6 to 5\%. It is generally better in the barrel region than in the endcaps, and also depends on the bremsstrahlung energy emitted by the electron as it traverses the material in front of the ECAL~\cite{CMS:2020uim,CMS-DP-2020-021}.

Events of interest are selected using a two-tiered trigger system. The first level (L1), composed of custom hardware processors, uses information from the calorimeters and muon detectors to select events at a rate of around 100\unit{kHz} within a fixed latency of 4\mus~\cite{CMS:2020cmk}. The second level, known as the high-level trigger, consists of a farm of processors running a version of the full event reconstruction software optimized for fast processing, and reduces the event rate to a few 1\unit{kHz} before data storage~\cite{CMS:2016ngn}.

\section{Simulated events}

To study the signature of highly boosted dielectron pairs originating from potential BSM processes, 
signal events are generated using the \PYTHIA 8.240 Monte Carlo (MC) event generator~\cite{pythia8} with benchmark models $\Pp\Pp\to\PX\to\PY\PY\to4\Pe$, where \PX and \PY are bosons with $\MX = 250$, 750, 2000\GeV and $\MY = 1$, 10\GeV. 
Physics processes containing isolated electrons are used for backgrounds. The \PW boson production, Drell-Yan (DY), and $\PQt\PAQt$ pair production processes are generated using the \MGvATNLO v2.6.5 MC event generator~\cite{madgraph}. A $\PBpm\to\PJGy\PKpm\to\EE\PKpm$ MC sample is generated using \PYTHIA, interfaced with the \EVTGEN v1.6.0~\cite{evtgen} particle decay simulation package. Lastly, $\PZ\to\MM\PGg$ events are generated using the \POWHEG v2.0 framework with the MiNNLO parton shower scheme~\cite{powhegMinnlo1,powhegMinnlo2}. The radiative correction for photons is addressed by \PHOTOS v215.5~\cite{photos}.

All event generators are interfaced with \PYTHIA with the CP5 underlying event tune~\cite{cpxTunes} to describe parton showering and hadronization. The NNPDF3.1 set of parton distribution functions is used~\cite{nnpdf}. The detector response is simulated using the \GEANTfour~\cite{geant4} package. The effect of additional proton-proton ($\Pp\Pp$) interactions in the same or adjacent bunch crossings (pileup) is incorporated into the simulation by including additional $\Pp\Pp$ interactions that have been generated with \PYTHIA. The simulated samples are reweighted to reproduce the pileup distribution equivalent to that of the mean number of interactions in data, as measured using the instantaneous luminosity per bunch crossing~\cite{LUM-17-003,CMS:2020ebo}.

\section{Merged electron ID}
In this section, we discuss the standard PF-based electron reconstruction algorithm and its \PeME ID deficiencies. The novel \PeME ID is based on electrons reconstructed by the default algorithm, but with additional requirements needed for \PeME ID. These selections must be independent of the standard electron ID to provide consistent efficiency across different environments where the signal \PeME is well isolated and where the \PeME ID efficiency is measured, as described in Section~\ref{sec:mergedEleID}. The modifications discussed are the association of a potential secondary track with the EB cluster, modifications of the cluster-track matching, and the use of the boosted decision tree (BDT) method that gives the optimal performance.

\subsection{Clustering and primary track selection}

{\tolerance=800 The standard EB electron reconstruction is initiated by clustering energy deposits in the ECAL. The clusters are grouped into a supercluster (SC)~\cite{CMS:2020uim} to incorporate bremsstrahlung photons. Because of the curvature of the electron trajectory in the solenoidal magnetic field, the bremsstrahlung photons increase the size of the cluster in the $\phi$ direction. Instead of using KF tracks (the default for other particles), electron tracks are reconstructed using an alternative method based on the GSF algorithm. The KF algorithm is a linear least squares estimator, optimal only if the energy loss in the matter is well described by a single Gaussian function of the energy fraction, $\zeta$, retained by the electron after traversing the material. Electron energy losses, however, are dominated by bremsstrahlung production. They follow the Bethe-Heitler distribution, which is highly asymmetric with a singularity at $\zeta\to1$ and a long tail toward $\zeta=0$. The GSF algorithm describes the probability density function as a sum of multiple Gaussian functions~\cite{Adam_2005}.\par}

Three or four consecutive hits in the pixel detector compatible with the SC position are selected to seed the GSF tracking step. The reconstructed GSF track is associated with the SC using geometrical matching to form an electron candidate~\cite{CMS:2020uim}. All electron candidates in the EB are considered inputs to the \PeME ID.
The associated GSF track is considered the primary track for the \PeME ID.

The standard electron reconstruction algorithm is suboptimal for \PeME for several reasons.
If a pair of electrons is as collimated as $\DR\equiv\sqrt{\smash[b]{(\Delta\eta)^{2}+(\Delta\phi)^{2}}} < 0.1$, which is smaller than 13\unit{cm} at the center of EB and thus smaller than six times the Moli{\`e}re radius of the lead tungstate crystal, the energy deposit of the two electrons may overlap.
If the dielectron is even more boosted, then the SC may fail to resolve the individual energy deposits, affecting the reconstruction of both electron objects and their associated GSF tracks. The GSF algorithm successfully builds two individual GSF tracks in 85\% of cases even when the collimation is as small as $\DR=0.01$. However, the shift in the SC's center of gravity (CoG) may cause GSF tracking failure for about 5\% of such clusters, which can be recovered by using KF tracks.

As the track of a single charged particle can be reconstructed more than once, the standard track reconstruction has a trajectory ``cleaner,'' which is applied after each iteration of the track finding sequence. When the fraction of shared hits is $>$19\%, the trajectory cleaner removes the track with fewer hits or larger $\chi^2$ if the number of hits is equal~\cite{CMS:2014pgm}. Therefore, if a pair of electrons is extremely boosted with $\gammaL > 200$, one of the two tracks may be removed by the trajectory cleaner. To recover these events, a single-track \PeME that does not require a secondary track is considered, and is discussed in more detail in Section~\ref{sec:BDTinputVariables}.

\subsection{Secondary track selection\label{sec:secondaryTrack}}

A key variable for distinguishing isolated electrons from \PeME's is the presence of a secondary track.
All tracks, either GSF or KF, within $\DR < 0.4$ of the electron's GSF track and passing the following track requirements are considered as secondary candidates. For KF tracks, if they are associated with a PF charged hadron,
the candidate is required to have its HCAL energy divided by its ECAL energy (\HoverE) less than 0.1. All secondary track candidates are required to pass quality criteria: $\pt > 10\GeV$, the difference between the track origin along the beam axis with that of the primary electron $\Delta z < 0.1\cm$, and the number of valid hits $\geq$8, with at least 1 in the pixel detector. KF tracks are also required to pass the standard high-quality track ID criteria (``high-purity'' tracks)~\cite{CMS:2014pgm}.
These requirements are listed in Table~\ref{tab:trackSelection}.
Because most electron tracks are reconstructed by both the KF and GSF algorithms, any KF track lying within $\DR < 0.001$ of the GSF primary-electron track is discarded to avoid double counting. Finally, GSF tracks are always preferred over KF tracks when multiple secondary track candidates exist. In addition, vertex fitting based on the KF algorithm~\cite{CMS:2014pgm} is performed by minimizing the mean square error estimated using the primary electron's track and the secondary track candidate. The $\chi^{2}$ probability of the fit is required to be $>$1\%. If ambiguity remains even within the same type of track collection, the secondary track with a higher vertex fit probability is selected. 
In the case of background with isolated electrons, secondary tracks passing these criteria mainly arise from bremsstrahlung photons that convert to an \EE pair inside the beam pipe or the first two layers of the pixel detector.

\begin{table*}[htb]
\centering
\topcaption{Secondary track selection criteria\label{tab:trackSelection}}
\begin{scotch}{lcc}
& GSF track & KF track \\
\hline
\pt & \multicolumn{2}{c}{$>$10\GeV}\\
$\Delta z$ with the primary track & \multicolumn{2}{c}{$<$0.1\cm}\\
Number of valid hits & \multicolumn{2}{c}{$\ge$8}\\
Number of valid pixel hits & \multicolumn{2}{c}{$\ge$1}\\
Vertex fit probability with the primary track & \multicolumn{2}{c}{$>$1\%}\\
\DR & $<$0.4 & $0.001 < \DR < 0.4$\\
Track quality & \NA & High purity\\
\HoverE & \NA & $<$0.1
\end{scotch}
\end{table*}

\subsection{Modified track-cluster matching}

\begin{figure}[hbtp]
  \centering
    \includegraphics[width=0.49\textwidth]{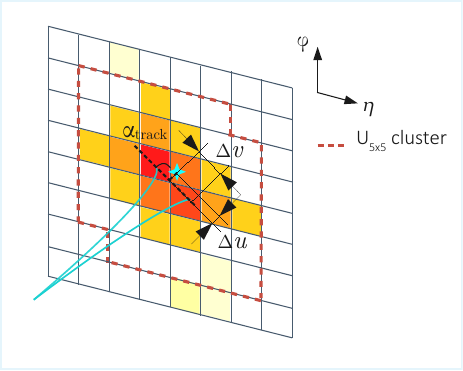}
    \caption{Visual representations of the variables \alphaTrack, \dUInfxf and \dVInfxf. Cyan-colored lines depict the incoming tracks of the dielectron. The black dashed line is used to define the $u$ and $v$ directions. The red dashed line represents the \ufxf cluster around the closest crystal from the tracks. The cyan-colored star is the log-weighted CoG of the \ufxf cluster.}
    \label{fig:union5x5Diagram}
\end{figure}

Geometrical matching between the SC and the associated track is a key discriminator between isolated electrons and merged dielectrons.
The standard electron ID uses the $\Delta\eta$ between the seed cluster's CoG and the GSF track's position on the EB surface (\dEtaInSeed), where the subscript ``in'' represents that the track is extrapolated from the innermost track position to the calorimeter surface. Similarly, the $\Delta\phi$ between the SC's CoG and the GSF track's position on the EB surface (\dXIn{\phi}) is also used. These can yield suboptimal ID performance for \PeME, since the overlapping dielectron's energy deposit can displace the CoG. New matching variables that are less sensitive to these effects are devised: \dUInfxf, \dVInfxf, \covfxfIeIe, \alphaTrack, and \alphafxf. The variables are defined as shown in Fig.~\ref{fig:union5x5Diagram}. The variable \alphaTrack denotes the angle between the $\phi$ axis and the line connecting the extrapolated position of two tracks to the EB surface. Then, a union of the $5\times5$ matrices of EB crystals (\ufxf) is defined around the closest crystal to the tracks, and its log-weighted CoG is calculated as a weighted average position (\varCoGufxf{x}) using Eq.~(\ref{eq:cog}),
 \begin{align}
    \varI{w} &= \max(0, 4.2+\ln(\varI{E}/\Eufxf)), \label{eq:wgt} \\
    \varCoGufxf{x}&=\frac{\sum_{i}^{\ufxf}\varI{w}\varI{x}}{\sum_{i}^{\ufxf}\varI{w}}. \label{eq:cog}
\end{align} 
A weight (\varI{w}) is defined as Eq.~(\ref{eq:wgt}), where \varI{x} and \varI{E} represent the position and energy of each crystal within the \ufxf cluster, and \Eufxf is the total energy. The weight is chosen to suppress the ECAL noise below 1.5\% of \Eufxf, and to have the same CoG if the set of crystals coincides for the \ufxf cluster and the standard SC. The deviation of this CoG with respect to the electron candidate's track represents the quality of track-cluster matching. The variable \dUInfxf (\dVInfxf) is defined as an angular separation perpendicular (parallel) to the line defined above. The variable \alphafxf is defined as the angle between the $\phi$ axis and the major axis of the log-weighted distribution of crystal energies calculated with the \ufxf cluster. The \ufxf cluster's covariance matrix is expressed by Eq.~(\ref{eq:cov}), \begin{gather}
\covfxfIxIx{a}{b} = \frac{\sum_{i}^{\ufxf}\varI{w}(\varI{a}-\varCoGufxf{a})(\varI{b}-\varCoGufxf{b})}{\sum_{i}^{\ufxf}\varI{w}}, \label{eq:cov}
\end{gather} where $a$ and $b$ can be either $\eta$ or $\phi$ axis, and $\varI{a} - \varCoGufxf{a}$ represents the distance to the $i$th crystal from the \ufxf cluster's CoG in units of a crystal.

\subsection{The BDT training\label{sec:BDTinputVariables}}

The final \PeME ID is based on two BDTs: for two-track and single-track candidates.
Table~\ref{tab:BDTvariablesHas2ndTrk} (\ref{tab:BDTvariablesNoTrk}) lists the variables used to train the two-track (single-track) model. The two-track model mainly uses geometrical matching between two tracks and the \ufxf cluster's CoG. The key variable for the single-track model is the ratio between the SC energy and the track momentum.
Geometrical matching variables between the cluster and the primary track also contribute to a lesser extent.

\begin{table*}[htb]
\centering
\topcaption{List of variables used to train the two-track model\label{tab:BDTvariablesHas2ndTrk}}
\renewcommand{\arraystretch}{1.2}
\begin{scotch}{cl}
Variable & Description \\
\hline
$\dVInfxf/\DR$ & \dVInfxf divided by $\DR$ between two tracks\\
\alphaTrack & Angle between the line connecting two tracks and the $\phi$ axis\\
\alphafxf & Angle between the major axis of the \ufxf cluster and the $\phi$ axis\\
\dXInUfxf{\phi} & $\Delta\phi$ between the extrapolated primary track and the CoG of \ufxf\\
\dXInUfxf{\eta} & $\Delta\eta$ between the extrapolated primary track and the CoG of \ufxf\\[\cmsTabSkip]
\multirow{2}{*}{\covfxfIeIe} & The log-weighted variance of distances between crystals of \ufxf\\
& along the $\eta$ axis (in units of number of crystals)
\end{scotch}
\end{table*}

\begin{table*}[htb]
\centering
\topcaption{List of variables used to train the single-track model\label{tab:BDTvariablesNoTrk}}
\renewcommand{\arraystretch}{1.2}
\begin{scotch}{cl}
Variable & Description \\
\hline
\EoverP & The SC energy divided by the GSF track's momentum\\
\dEtaInSeed & $\Delta\eta$ between the seed cluster and the extrapolated GSF track\\
\dXIn{\phi} & $\Delta\phi$ between the SC and the extrapolated GSF track\\
$E_{1\times5}/E_{5\times5}$ & The energy ratio between the arrays of $1\times5$ and $5\times5$ crystals (in $\eta\times\phi$)\\[\cmsTabSkip]
\multirow{2}{*}{\covIxIx{\eta}{\eta}{\mathrm{SC}}} & The log-weighted variance of distances between crystals of SC\\
& along the $\eta$ axis (in units of number of crystals)
\end{scotch}
\end{table*}

The BDTs are trained using merged clusters from $\PX\to\PY\PY\to4\Pe$ events as signals, and isolated
electrons from \PW boson production, the DY process, and $\PQt\PAQt$ pair production as background.
Preselection criteria are applied. For the $\PX\to\PY\PY\to4\Pe$ signal hypothesis, at least two reconstructed \PeME's are expected in the event after merging. Accordingly, two electron objects with $\pt > 20\GeV$ are required. Both single-track and two-track \PeME candidates are required to match a pair of generated electrons within $\DR < 0.3$. For the two-track case, the primary and secondary tracks must be located within $\DR < 0.001$ of the generated electrons to avoid ambiguity in track matching.

The single-track case mainly occurs when the dielectron has sufficiently high \pt to have a small enough opening angle that the trajectory cleaning feature is used. Therefore, a minimum \pt requirement of 200\GeV is applied to prevent the model from learning the correlation between \pt and other input variables. Also, electrons are required to have $\Delta\pt/\pt < 0.5$, where $\Delta\pt$ represents the uncertainty in the track's \pt measurement, to avoid learning features reflecting the poorer track momentum description at high energy. 

\begin{figure*}[hbtp]
  \centering
    \includegraphics[width=0.45\textwidth]{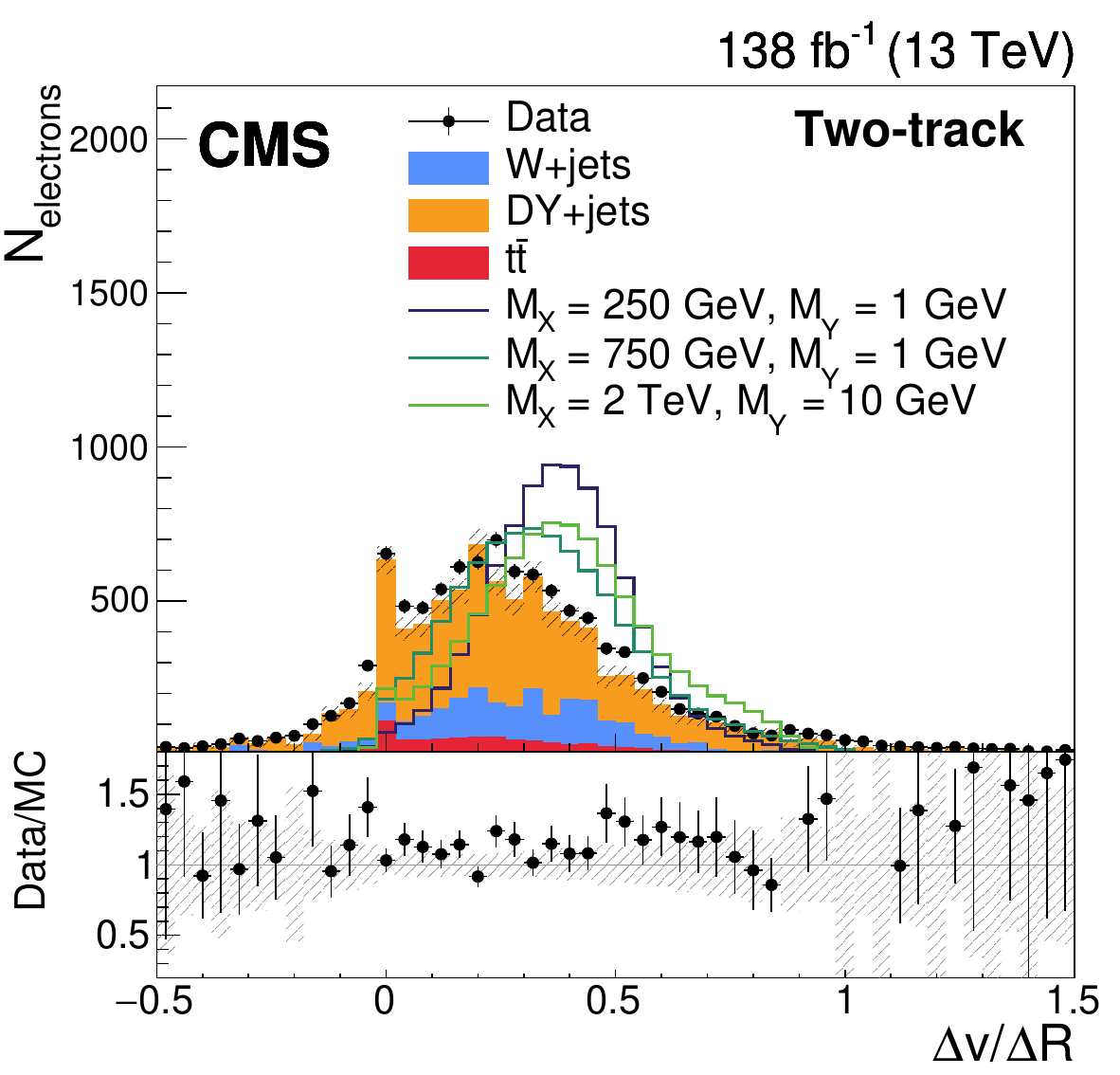}
    \includegraphics[width=0.45\textwidth]{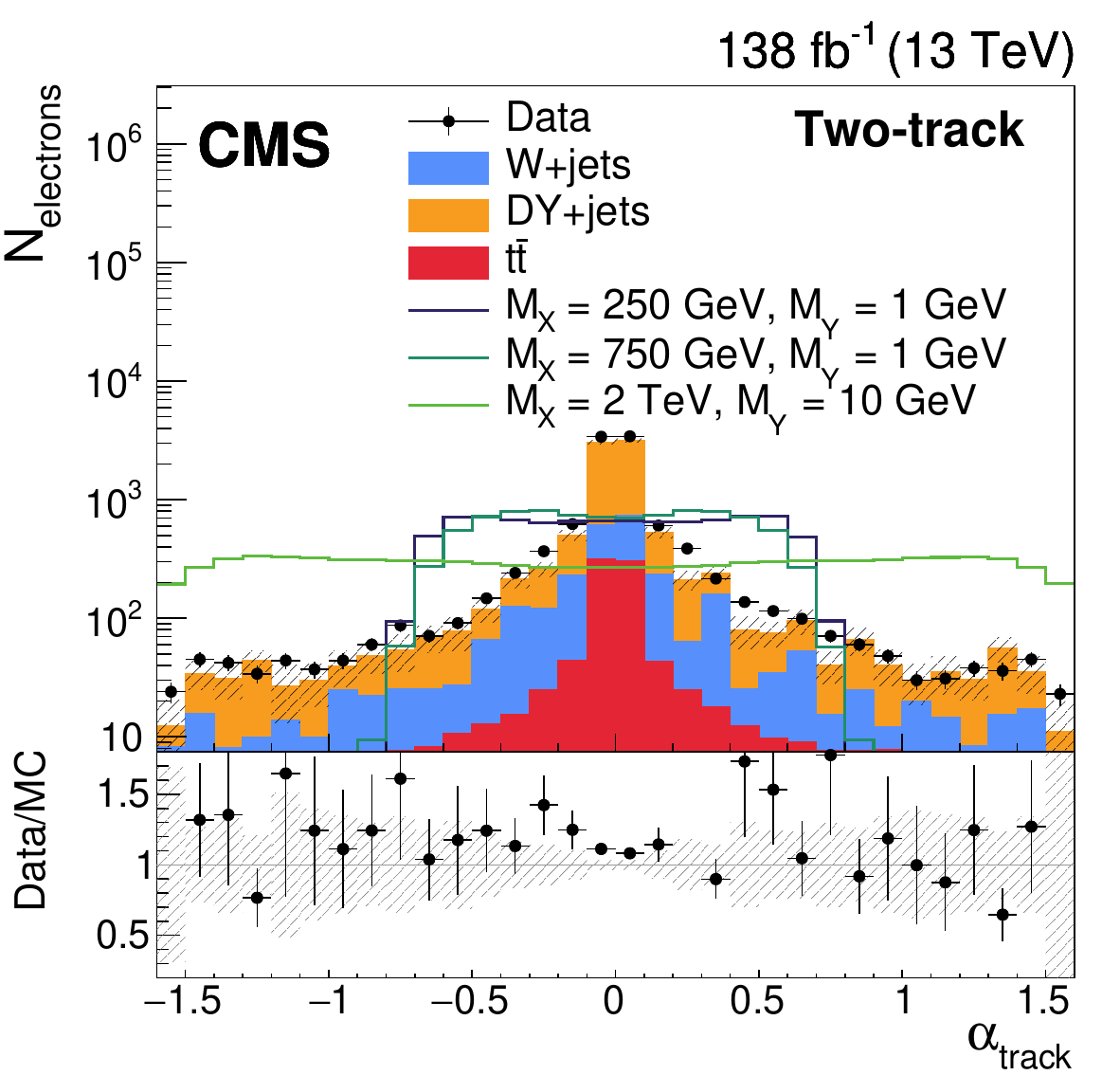}\\
    \includegraphics[width=0.45\textwidth]{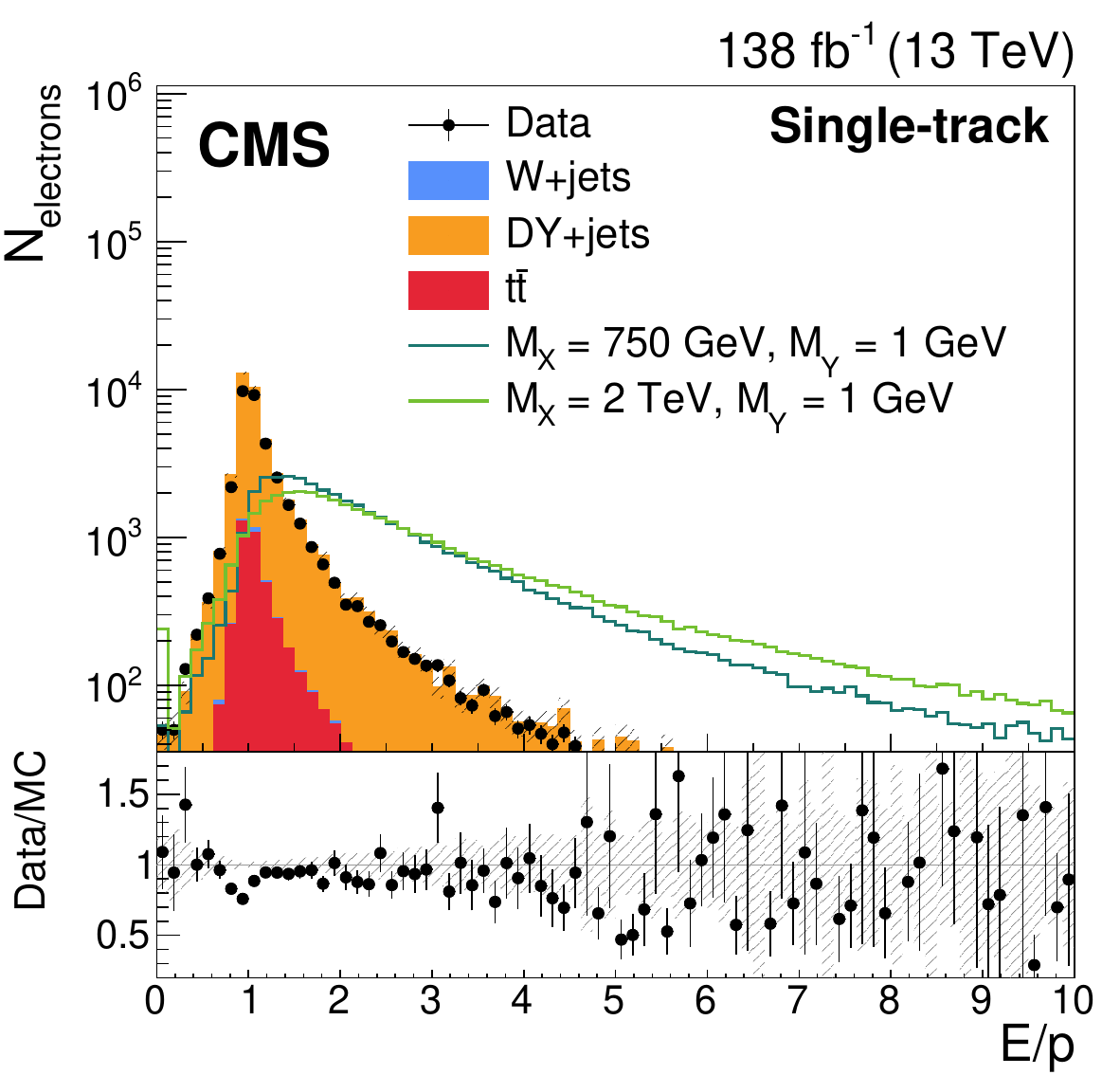}
    \includegraphics[width=0.45\textwidth]{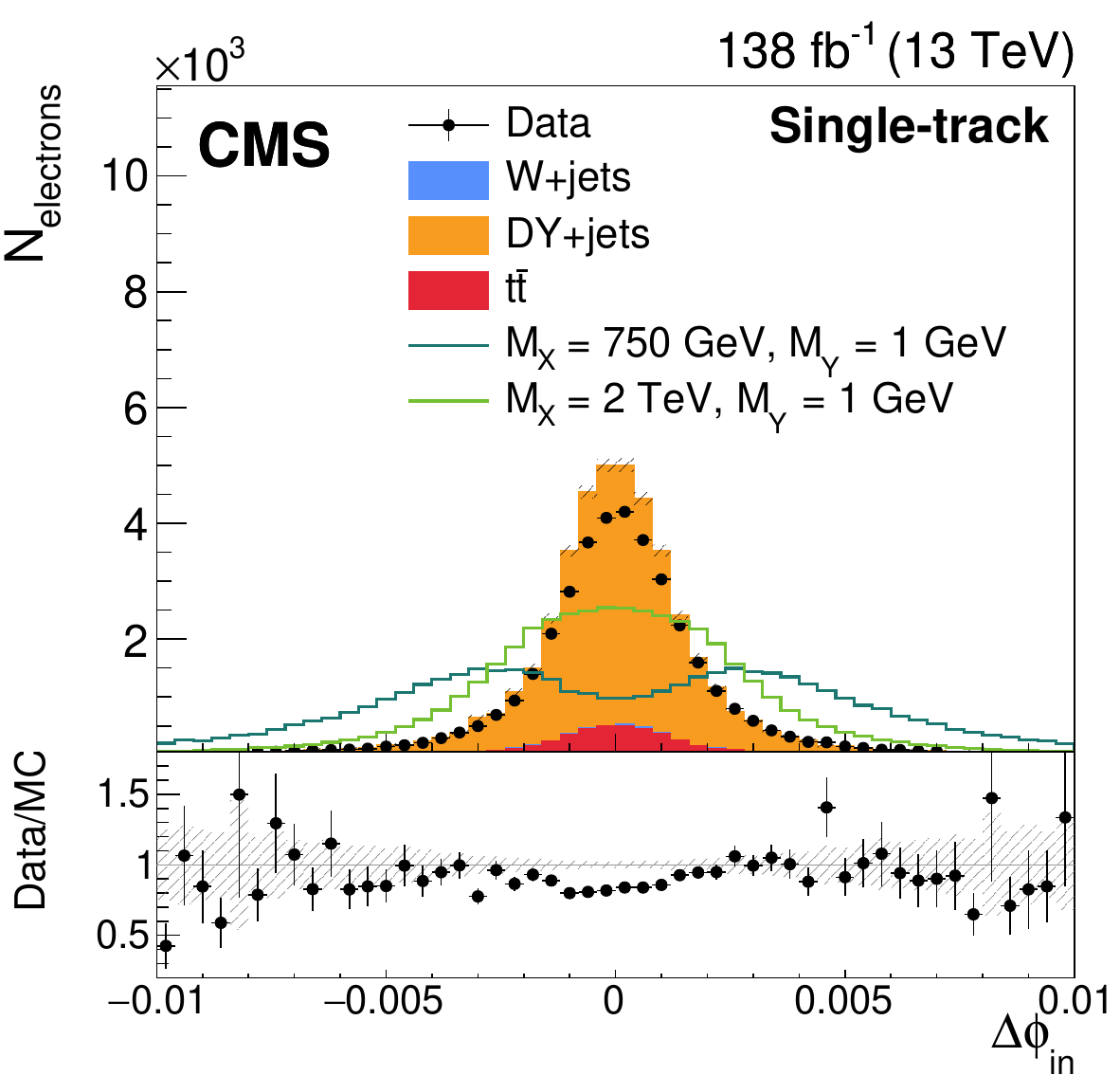}
    \caption{The distribution of the two most contributing variables for the (upper) two-track and (lower) single-track models: (upper left) $\dVInfxf/\DR$, (upper right) \alphaTrack, (lower left) \EoverP, and (lower right) \dXIn{\phi}. The signal samples are displayed according to the resonance masses that match the model. The vertical bar and shaded band represent the statistical uncertainty of the data and MC simulation, respectively. Each lined signal histogram is normalized to the background yield.}
    \label{fig:BDTinputVariables}
\end{figure*}

Figure~\ref{fig:BDTinputVariables} shows distributions of several key ID variables for signal and background events. Data collected with a trigger that requires two SCs with $\HoverE < 0.15$ and transverse energy (\ET) greater than 60\GeV (2016) or 70\GeV (2017 and 2018) are used, corresponding to an integrated luminosity of 138\fbinv~\cite{LUM-17-003,LUM-17-004,LUM-18-002}. The \ET is defined as the projection of the vector onto the plane perpendicular to the beam axis, where the vector points to the SC from the interaction point with the magnitude equivalent to the SC's energy. The trigger requirement has no shower shape or track-cluster matching criteria, and it does not bias the ID variable distributions. Corresponding electron \pt thresholds are applied for the comparison between data and MC simulation. For \PeME, the \ufxf cluster's CoG should be located between the two tracks, yielding a $\dVInfxf/\DR$ value closer to $0.5$. However, single electrons can have a peak close to zero if they have an accidental secondary track that does not deposit energy in the EB. Variables such as \alphaTrack and \alphafxf also contribute to the discrimination, since the decay of a massive particle would produce nonzero recoil momenta along the $\eta$ axis. The backgrounds, such as converted bremsstrahlung photons, align along the $\phi$ axis because of the magnetic field. Also, the \EoverP of an \PeME has a higher value, while the background peaks at unity. Lastly, \dXIn{\phi} can contribute to the ID even if there is no secondary track because the oppositely charged electron can still shift the SC's CoG in the $\phi$ direction.

\begin{figure}[hbtp]
  \centering
    \includegraphics[width=0.49\textwidth]{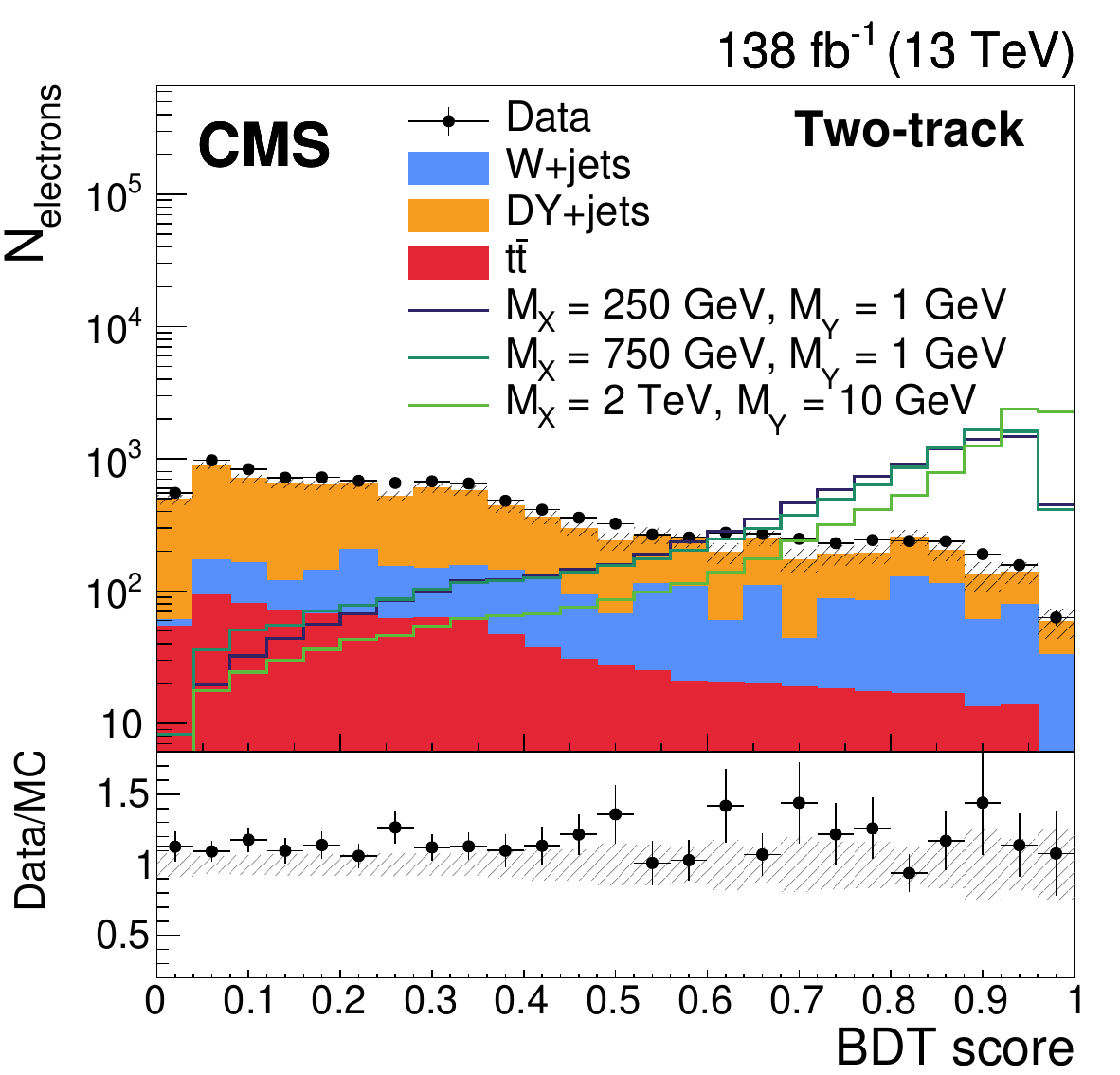}
    \includegraphics[width=0.49\textwidth]{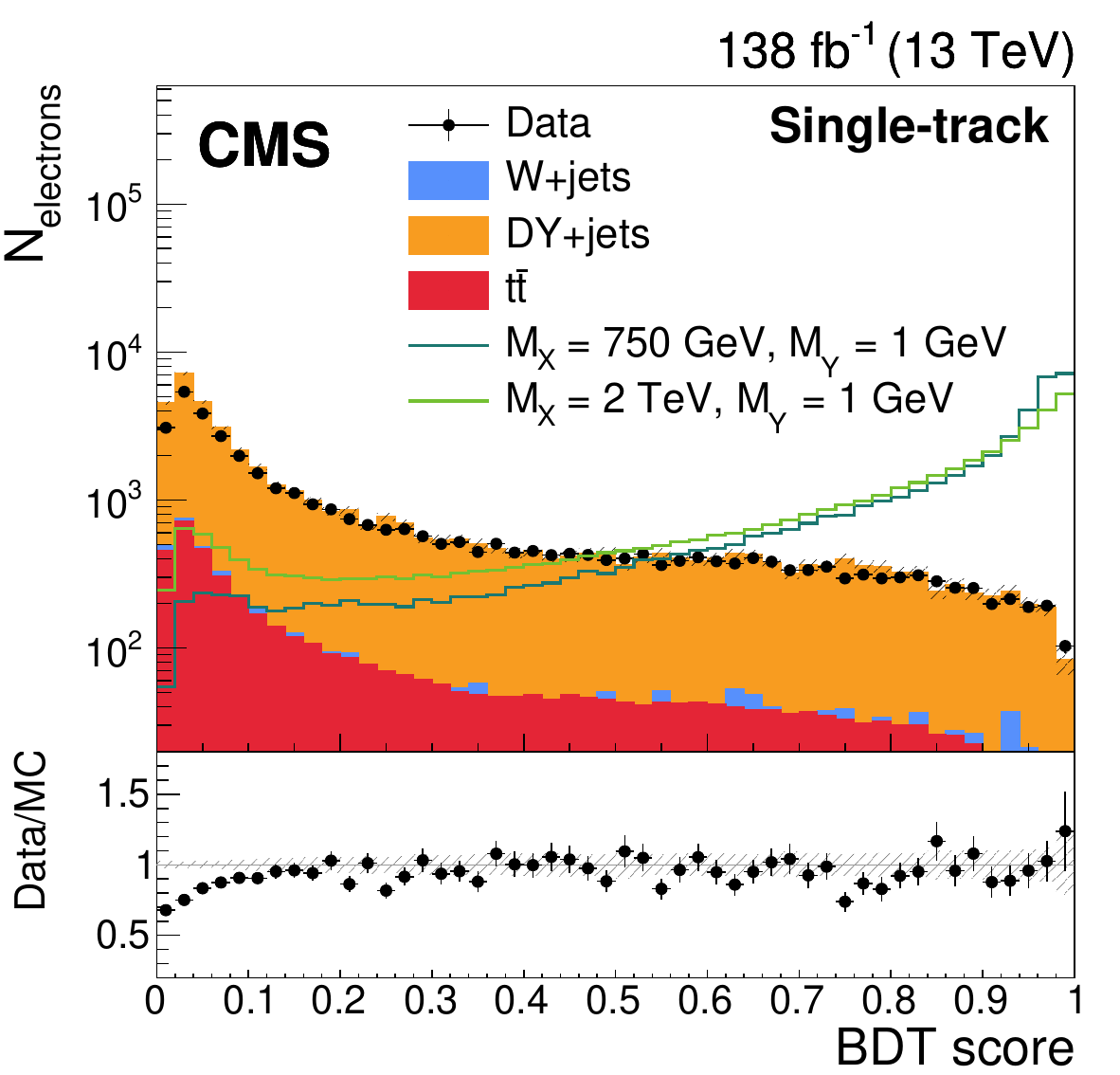}
    \caption{The BDT score distributions of the (\cmsLeft) two-track and (\cmsRight) single-track models. The signal samples are displayed according to the resonance masses that match the model. The vertical bar and shaded band represent the statistical uncertainty of the data and MC simulation, respectively. Each lined signal histogram is normalized to the background yield.}
    \label{fig:BDTscore}
\end{figure}

\begin{figure}[hbtp]
  \centering
    \includegraphics[width=\cmsFigWidth]{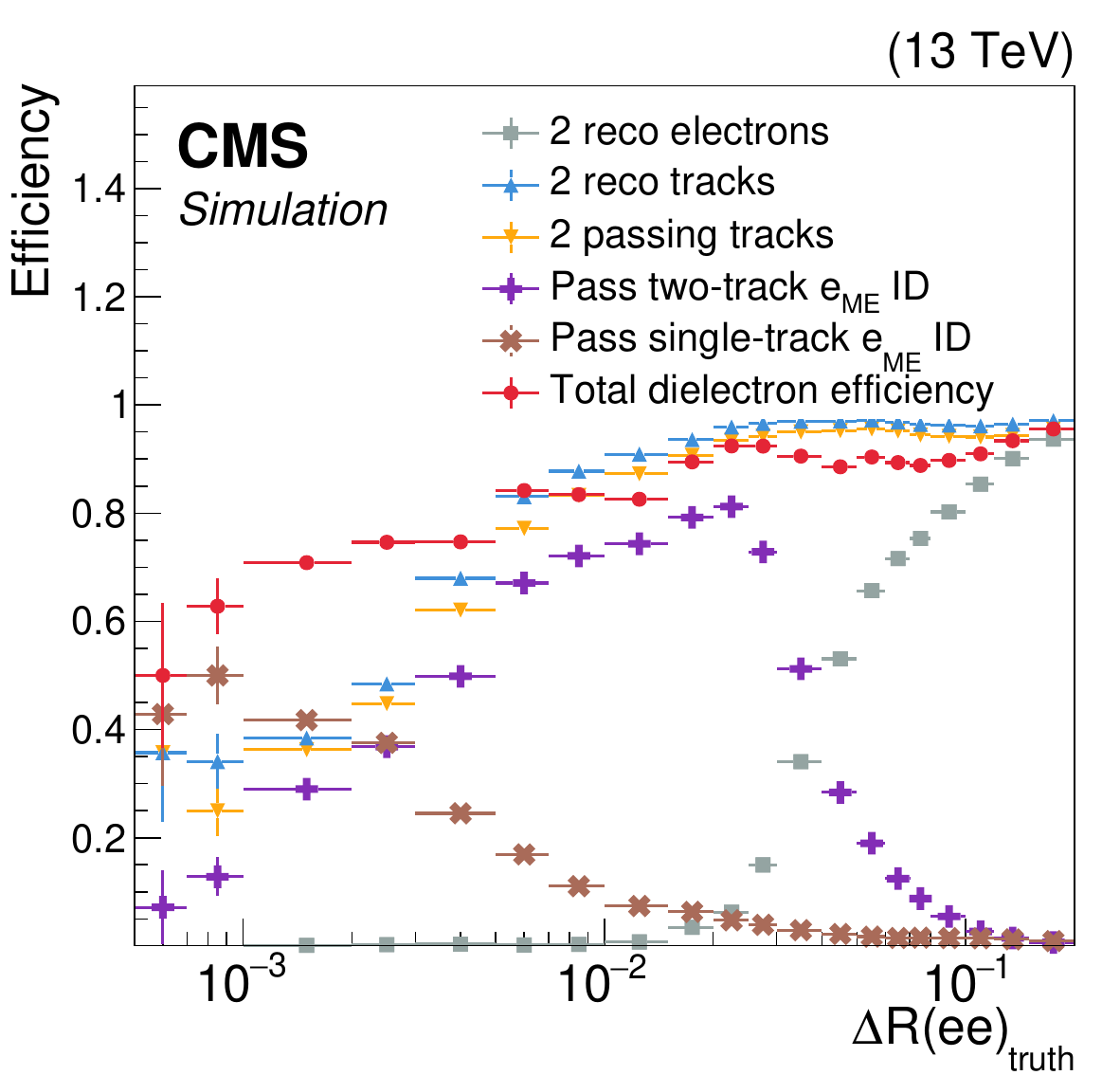}
    \caption{The signal dielectron selection efficiency as a function of \DR. The vertical error bar represents the statistical uncertainty. An equal number of signal events with $\MX = 250$, 750, 2000\GeV and $\MY = 1$, 10\GeV are used for each mass point to incorporate dielectrons with various \DR. The ID efficiencies include the effects of all prerequisite selections. For instance, the \PeME ID efficiency with two tracks (purple) includes the efficiency of the track selection described in Table~\ref{tab:trackSelection} (yellow). The total dielectron efficiency is a sum of the two-track (purple) and single-track (brown) \PeME ID efficiencies, and the standard reconstruction efficiency of two electrons (gray).}
    \label{fig:finalIDefficiency}
\end{figure}

The BDT models are trained using \textsc{XGBoost}~\cite{xgboost} and the model hyperparameters are optimized using \textsc{Hyperopt}~\cite{hyperopt}. Simulated events are split into five folds: one is used for testing, and the others for training. The training is repeated for each possible combination of sets, and the model with the largest area under the curve of signal efficiency versus background efficiency is selected. For the two-track model, electrons were reweighted so that the distribution of the \ET of their \ufxf cluster (\ETfxf) is uniform to avoid potential biases due to the different kinematics between background and signal. Figure~\ref{fig:BDTscore} shows the BDT score distribution for the two-track and single-track models. The threshold on the BDT score is chosen to retain 10\% background efficiency, and is approximately 0.6 for both models.
Figure~\ref{fig:finalIDefficiency} shows the dielectron signal ID efficiency as a function of \DR between the two electrons at several cumulative levels: passing the electron object reconstruction, the secondary track selection, and the \PeME ID. The denominator is the number of MC truth dielectrons. The numerator for \PeME ID efficiencies requires passing the ID and not being reconstructed as two resolved electrons. The total dielectron efficiency includes the cases where the dielectron is reconstructed as two objects by the standard reconstruction algorithm and where the pair is reconstructed as a single object and passes the \PeME ID.

\section{Merged electron ID data and MC efficiencies\label{sec:mergedEleID}}

The \PeME ID efficiencies in data are measured for each model. The two-track model efficiency is measured using dielectrons from boosted \PJGy mesons. Events with $\PJGy\to\EE$ candidates are selected from the dataset collected by tagging \PQb hadron decays with muon triggers (``\PB Parking'')~\cite{Bparking} and looking for \PQb hadrons from the same $\Pp\Pp$ collision decaying to $\PJGy+\mathrm{X}$. However, $\PJGy\to\EE$ decays can only be distinguished from backgrounds by examining the invariant mass of the two electron tracks. Therefore, the single-track model efficiency is estimated using collimated dielectrons from photon conversions where only one of the electron tracks is reconstructed (``single-leg'' conversions). The single-leg conversions are identified based on the trajectories' impact parameters and the number of missing hits in the inner tracker~\cite{CMS:2020uim}.

\subsection{The two-track model efficiency\label{sec:JpsiControlRegion}}

\begin{figure}[hbtp]
  \centering
    \includegraphics[width=0.49\textwidth]{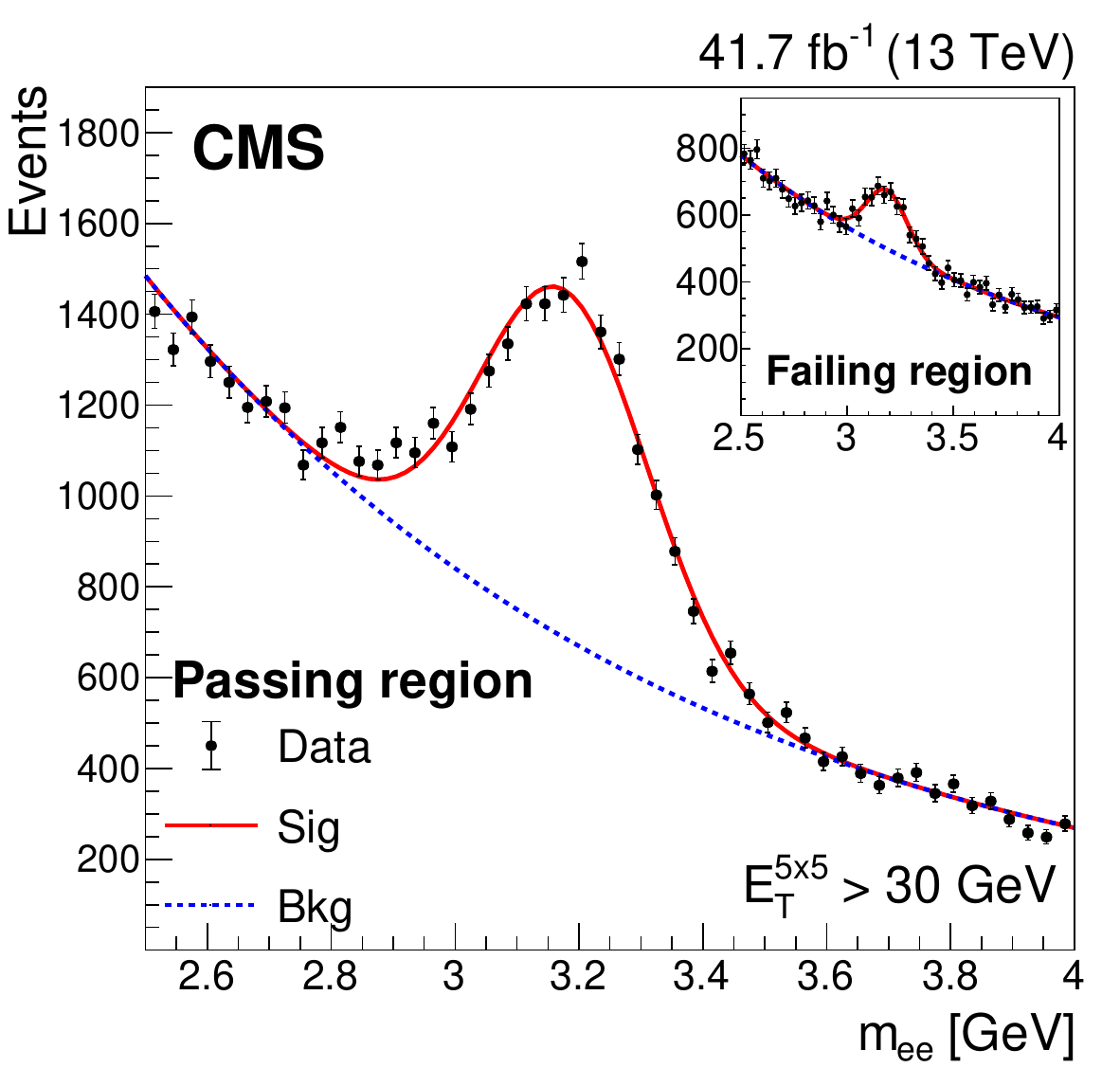}
    \caption{The nominal dielectron mass distribution of \PJGy candidates in the \PB Parking dataset with $\ETfxf > 30\GeV$ that pass or fail the two-track \PeME ID. The dielectron mass is reconstructed using the tracks of electron candidates.}
    \label{fig:JpsiMass_InvM}
\end{figure}

Candidate $\PJGy\to\EE$ decays are selected from the \PB Parking dataset, corresponding to an integrated luminosity of 41.7\fbinv. The events are triggered by a low-\pt muon displaced from the beam axis. The \EE pair originating from the same $\Pp\Pp$ collision as the trigger muon is considered a candidate $\PJGy\to\EE$ decay. Table~\ref{tab:JpsiCRselection} shows the selection criteria for the $\PJGy\to\EE$ control region. The trigger muon needs to satisfy $\pt > 9\GeV$, $\abs{\eta} < 1.5$, and the ratio between the transverse impact parameter in the plane perpendicular to the beam axis (\dxy) and its uncertainty must satisfy $\dxy/\Delta \dxy > 6$. Also, it needs to pass the standard ID criteria optimized for both prompt muons and muons from heavy-flavor decays with medium efficiency (``medium ID'')~\cite{CMS:2018rym}. The \ETfxf of the \PeME candidate must be $>$20\GeV. The $\Delta z$ between electron tracks and the trigger muon needs to be $<$0.1\cm, while the \DR between them needs to be $>$0.1 to disambiguate the muon and secondary track. The secondary track must pass the criteria described in Section~\ref{sec:secondaryTrack}. The number of $\PJGy\to\EE$ candidates passing or failing the \PeME ID criteria was estimated by fitting the resonance structure in the invariant mass of the two tracks in the range $2.5 < \Mee < 4\GeV$. Figure~\ref{fig:JpsiMass_InvM} shows the dielectron mass distribution of \PJGy candidates with $\ETfxf > 30\GeV$. A Gaussian function is used to describe the nominal signal shape. The nominal background shape is modeled by an exponential function. The \PeME ID efficiency is derived using the number of \PJGy candidates that pass the two-track \PeME ID divided by the total number of \PJGy candidates. The systematic uncertainty of the fit was evaluated by using alternative signal and background functions. A single-sided Crystal Ball function~\cite{Oreglia:1980cs} convoluted with an exponential tail at low masses is used as an alternative signal. The alternative background shape is an exponential decay at high masses and an error function at low masses.

\begin{table}[htb]
\centering
\topcaption{Selection criteria for the $\PJGy\to\EE$ control region.\label{tab:JpsiCRselection}}
\renewcommand{\arraystretch}{1.4}
\begin{scotch}{ll}
\multirow{3}{*}{Trigger muon} & $\pt > 9\GeV$\\
& $\abs{\eta} < 1.5$\\
& $\dxy / \Delta \dxy > 6$\\
\hline
\multirow{7}{*}{Electron} & $\ETfxf > 20\GeV$\\
& $\abs{\eta_{\mathrm{SC}}} < 1.4442$\\
\cline{2-2}
& $\Delta z(\Pe_{1}, \PGm_{\mathrm{tag}}) < 0.1\cm$\\
& $\Delta z(\Pe_{2}, \PGm_{\mathrm{tag}}) < 0.1\cm$\\
& $\DR(\Pe_{2}, \PGm_{\mathrm{tag}}) > 0.1$\\
\cline{2-2}
& Vertex fit probability $>$0.01\\
& $2.5 < \Mee < 4.0\GeV$\\
\end{scotch}
\end{table}

\begin{figure}[hbtp]
  \centering
    \includegraphics[width=0.49\textwidth]{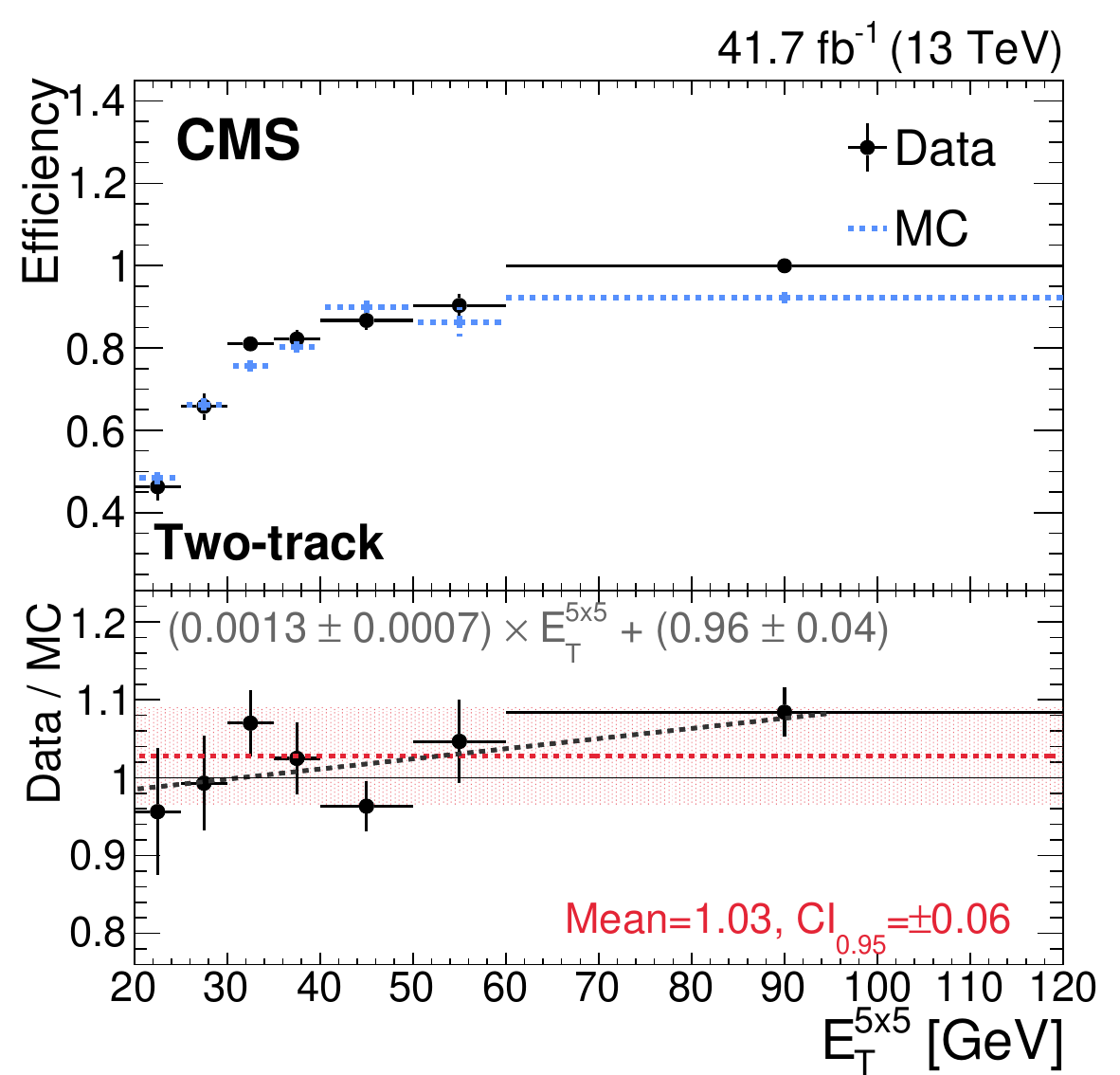}
    \includegraphics[width=0.49\textwidth]{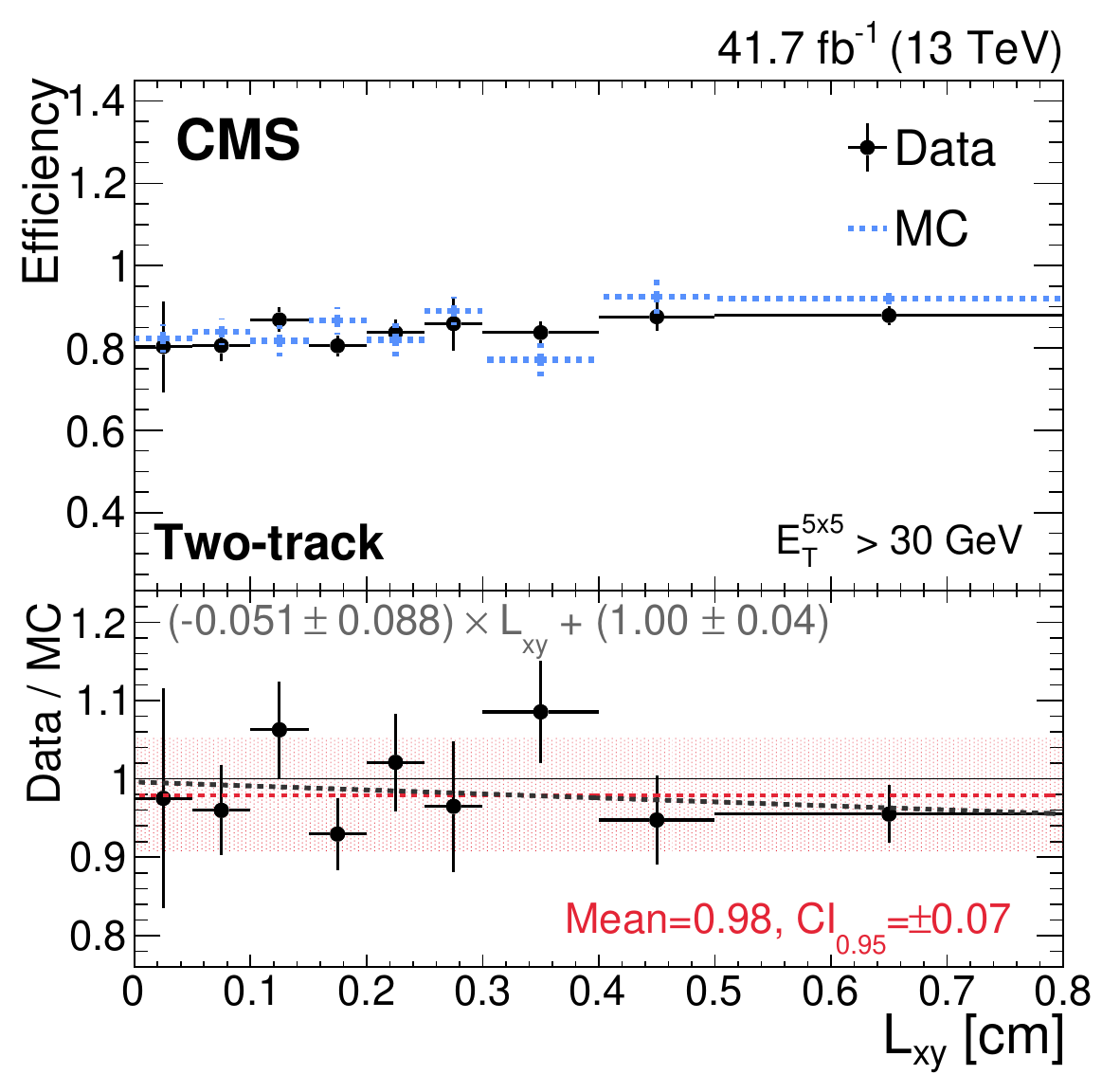}
    \caption{The efficiency and SF for the two-track model as a function of (\cmsLeft) \ETfxf and (\cmsRight) \Lxy with $\ETfxf > 30\GeV$. The vertical bar shows the statistical and systematic uncertainties from the alternative fits, summed in quadrature. The shaded band depicts the \CItwoSig. The red and gray dashed lines depict the constant and first-order polynomial fits, respectively.}
    \label{fig:JpsiMass_Has2ndTrk}
\end{figure}

Although boosted $\PJGy\to\EE$ candidates exhibit characteristics similar to the signal \PeME, there are also intrinsic differences because most \PJGy mesons are produced from \PQb hadron decays that are displaced from the collision point, and have relatively low momenta. Therefore, the \PeME ID efficiency is parametrized as a function of \ETfxf and the transverse distance (\Lxy) from the beam axis to the vertex. Figure~\ref{fig:JpsiMass_Has2ndTrk} shows the efficiency as a function of \ETfxf and \Lxy. The efficiency scale factor (SF) is defined as the ratio between the efficiencies in data and MC simulation, where the MC efficiency is estimated by counting \PJGy candidates. No evident dependence on \Lxy is observed. To take into account the different \ETfxf ranges between the boosted \PJGy candidates and various signal mass hypotheses, an \ETfxf-dependent systematic uncertainty in the SF is introduced. The nominal efficiency SF is about 1.03, estimated by averaging over all \ETfxf bins. The systematic uncertainty is approximately 6\% based on the 95\% confidence interval (\CItwoSig), on average. In addition, a fit assuming the SF is linear in \ETfxf is performed using a first-order polynomial, and the discrepancy with the nominal value is added to the uncertainty in quadrature. From the definition of the efficiency, the SF's systematic variations are not allowed to exceed unity divided by the MC efficiency. The additional uncertainty from the extrapolation covers any potential \ETfxf dependence of the SF.

\subsection{The single-track model efficiency}

The ID efficiency of the single-track model is measured using single-leg conversions. The final-state radiation (FSR) in $\PZ\to\MM\PGg$ events is an excellent source of high-purity photons thanks to the low-background environment of the muon final states and $\PZ\to\MM$ background suppression using an invariant mass requirement. Table~\ref{tab:convCRselection} shows the selection criteria for the single-leg conversion control region. The muons need to have $\pt > 10\GeV$, $\abs{\eta} < 2.4$, and pass the medium ID criteria. Isolated single-muon triggers that require $\pt > 24$ (27)\GeV and $\abs{\eta} < 2.4$ were available during the 2016 and 2018 (2017) data-taking periods, with the total integrated luminosity of 138\fbinv. The \pt of a muon that triggers the event must be 2\GeV above the trigger \pt threshold to be in the plateau region of the trigger efficiency curve~\cite{CMSmuonTrigger}. Also, the muon's tracker isolation must not exceed 10\% of the muon \pt, where the tracker isolation is defined as the scalar \pt sum of all tracks (except the muon's) from the same vertex within $\DR < 0.3$ around the muon. As the FSR photon is likely collinear with the muon, the $\DR$ between the muon that did not trigger the event and the FSR photon must be smaller than 0.8. However, the minimum $\DR$ between the photon and muons is kept at 0.05 to avoid potential effects due to the energy deposit in ECAL left by the muon. Lastly, the invariant mass of the two muons and the photon needs to fulfill $70 < \Mmmg < 110\GeV$ and $\Mmmg + \Mmm < 180\GeV$ to ensure that the photon originates from FSR.

\begin{table*}[htb]
\centering
\topcaption{Selection criteria for the $\PZ\to\MM\PGg$ control region.\label{tab:convCRselection}}
\renewcommand{\arraystretch}{1.4}
\begin{scotch}{lcc}
\multirow{6}{*}{Muons} & Trigger muon ($\PGm_{\mathrm{trig}}$) & FSR muon ($\PGm_{\mathrm{FSR}}$)\\
\cline{2-3}
& \multicolumn{1}{l}{$\pt > 26\GeV$ (2016, 2018)} & \multirow{2}{*}{$\pt > 10\GeV$}\\
& \multicolumn{1}{l}{$\pt > 29\GeV$ (2017)} & \\
& \multicolumn{2}{c}{$\abs{\eta} < 2.4$}\\
& $\sum^{\DR < 0.3}_{\mathrm{tracks}}\pt/{\pt}_{\PGm} < 0.1$ & \NA\\
\hline
\multirow{5}{*}{Converted photon} & \multicolumn{2}{c}{$\ETfxf > 10\GeV$}\\
& \multicolumn{2}{c}{$\abs{\eta_{\mathrm{SC}}} < 1.4442$}\\
& \multicolumn{2}{c}{$\DR(\PGm,\PGg) > 0.05$}\\
& \multicolumn{2}{c}{$\DR(\PGm_{\mathrm{FSR}},\PGg) < 0.8$}\\
& \multicolumn{2}{c}{$70 < \Mmmg < 110\GeV$ and $\Mmmg + \Mmm < 180\GeV$}\\
\end{scotch}
\end{table*}

\begin{figure}[hbtp]
  \centering
    \includegraphics[width=0.49\textwidth]{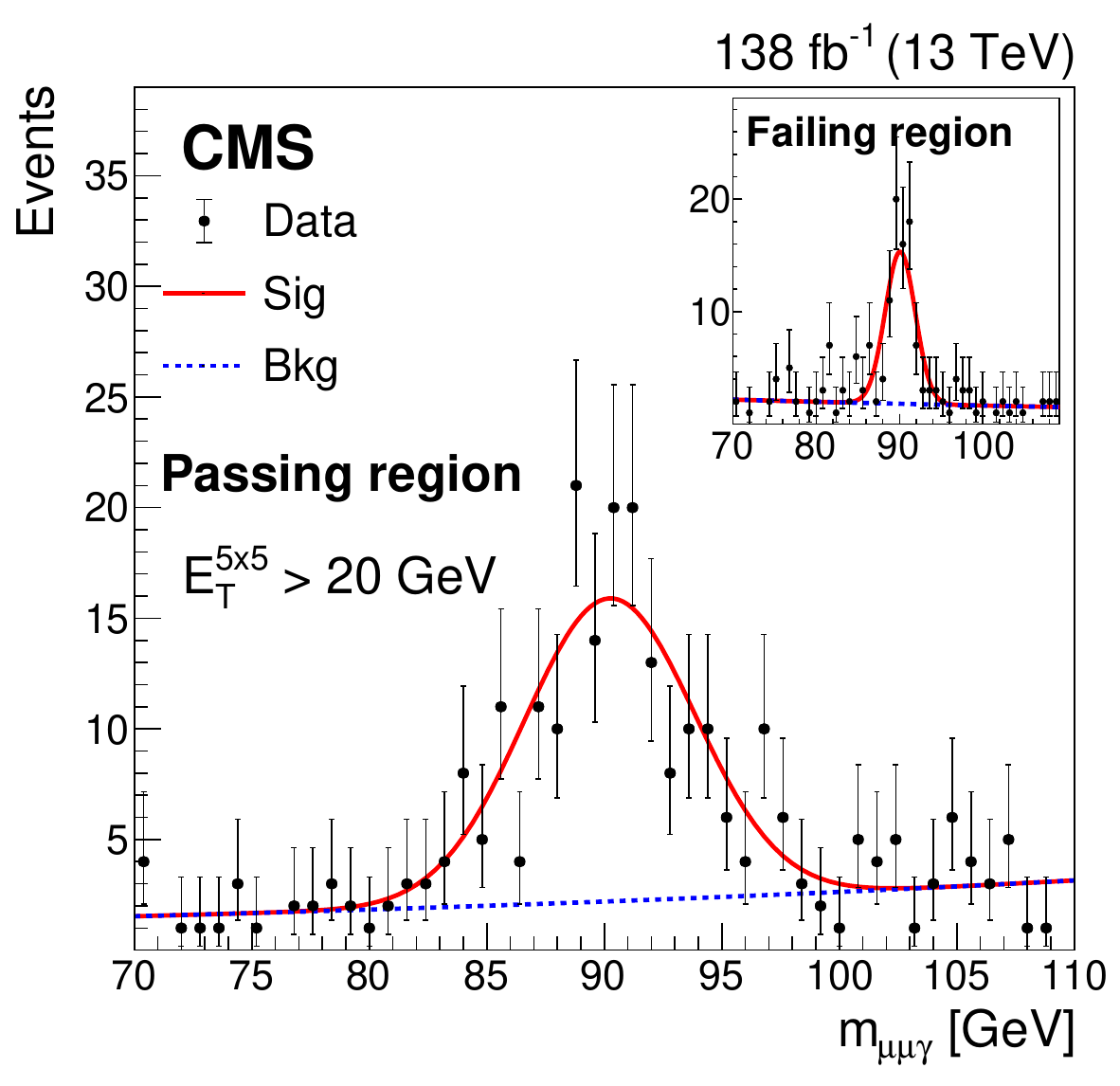}
    \caption{The nominal \PZ boson candidate mass distribution in data using $\MM\PGg$ events with $\ETfxf > 20\GeV$. The passing and failing regions represent the \PZ boson candidate mass distributions with \PeME candidates that pass or fail the \PeME ID.}
    \label{fig:mmgMass_InvM}
\end{figure}

\begin{figure}[hbtp]
  \centering
    \includegraphics[width=0.49\textwidth]{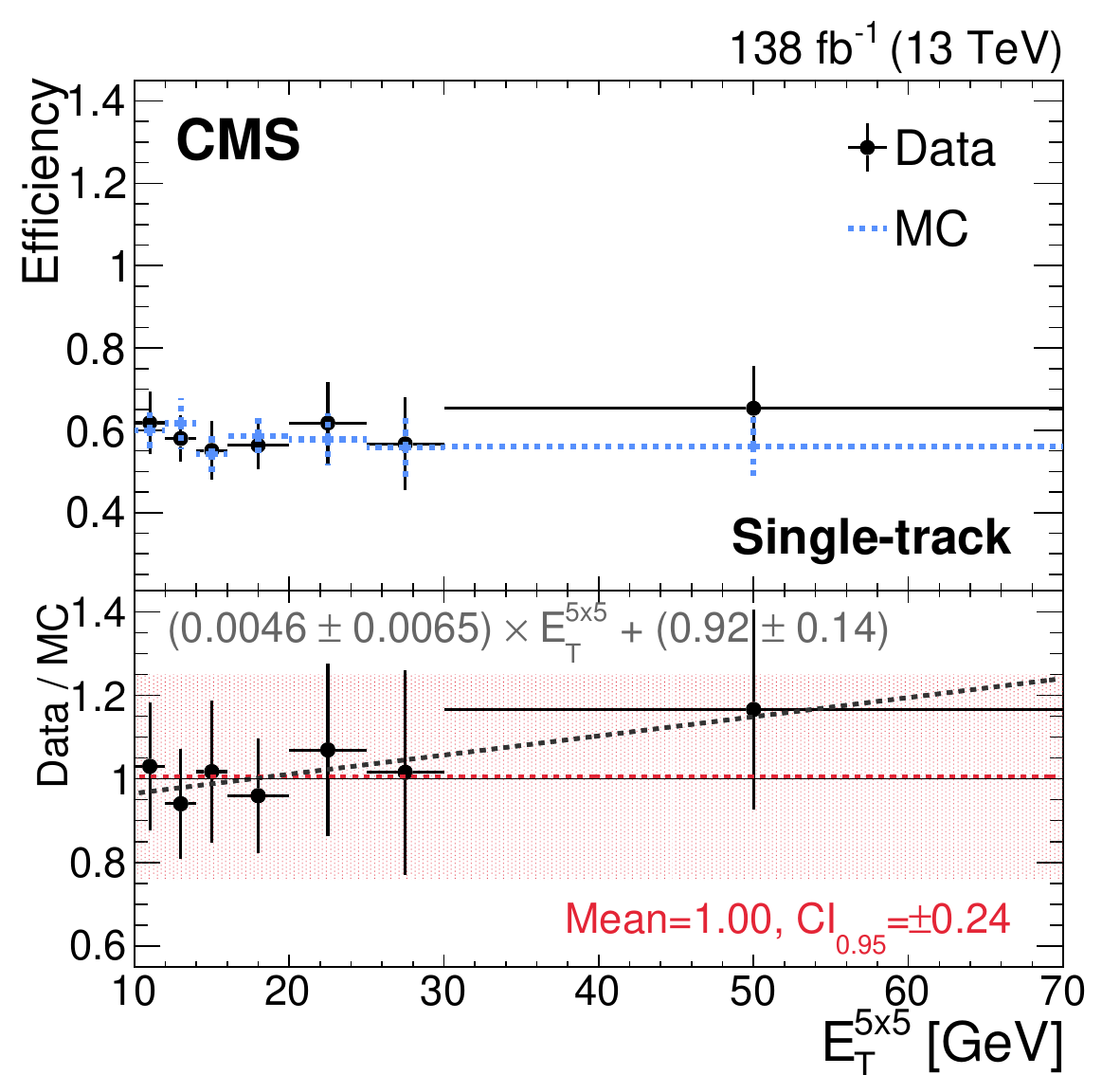}
    \includegraphics[width=0.49\textwidth]{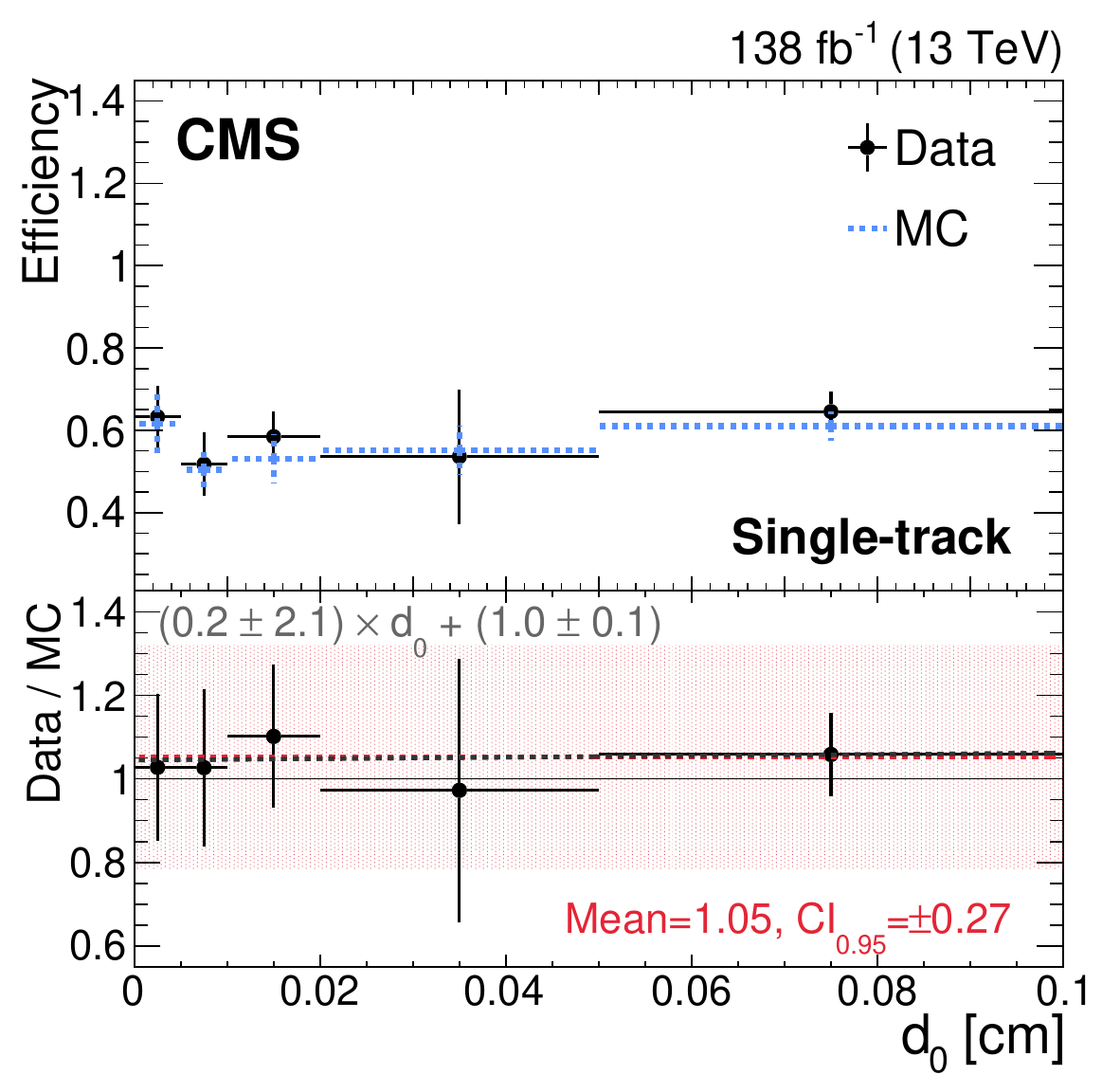}
    \caption{The efficiency and SF as a function of (\cmsLeft) \ETfxf and (\cmsRight) \dzero for the single-track model. The vertical bar shows the statistical and systematic uncertainties from the alternative fits, summed in quadrature. The shaded band depicts the \CItwoSig. The red and gray dashed lines depict the constant and first-order polynomial fits, respectively.}
    \label{fig:mmgMass_NoTrk}
\end{figure}

Figure~\ref{fig:mmgMass_InvM} shows the nominal \PZ boson candidate mass distribution of $\MM\PGg$ events in data where the \EE pair from the converted photon becomes the \PeME candidate. The choice of nominal and alternative signal and background shapes is the same as was used for the two-track model. Analogous to the case with secondary tracks, the efficiency is derived using the number of \PZ boson candidates that pass the single-track \PeME ID divided by the total number of \PZ candidates. Figure~\ref{fig:mmgMass_NoTrk} represents the efficiency as a function of \ETfxf and \dzero, where \dzero is the three-dimensional (3D) distance of closest approach (DCA) from the PV to the electron track. No apparent trend is seen in \dzero.

Similar to the case with secondary tracks, the final SF is taken from a constant fit of bin-by-bin SFs. The \CItwoSig of the fit is assigned as the uncertainty. The SF is found to be unity, but with about 24\% uncertainty. The difference between the constant and first-order polynomial fit is added to the uncertainty in quadrature, with an upper bound of unity divided by the MC efficiency. This yields a maximum total uncertainty of approximately 40\%. The effect of different topologies between the signal \PeME and the converted photon is also assessed. The efficiency in the signal MC sample ranges 65--75\% for the single-track case. The efficiency for the $\PZ\to\MM\PGg$ events is about 60\%. Considering that the observed discrepancy is a result of differences in both \ETfxf and event topology, and is already well covered by the systematic uncertainty, no additional uncertainty is taken into account.

\section{Merged electron energy scale and resolution}

The energy of standard electrons, estimated by combining the GSF track momentum and the corresponding SC energy, is dominated by the ECAL measurement for $\pt > 20\GeV$~\cite{CMS:2020uim}. The single-track \PeME is highly collimated, and the standard SC is sufficient to capture all energy deposits from the dielectron. For the two-track \PeME, however, the \DR separation between the electrons ranges from $\mathcal{O}(10^{-2})$ to $\mathcal{O}(10^{-1})$. In this case, the standard SC is not guaranteed to capture all energy deposits from the dielectron. Therefore, the \ufxf cluster's energy is used to describe the \PeME energy. The energy scale and resolution of the \ufxf cluster in data can be characterized using a resonance containing a boosted \EE pair as one of the decay products. The $\PBpm\to\PJGy\PKpm\to\EE\PKpm$ decay is an excellent candidate because it is one of the most common \PQb hadron decays containing $\PJGy\to\EE$. Also, the kaon from this decay typically has $1 < \pt < 10\GeV$, where the track momentum resolution is about 1\%~\cite{CMS:2014pgm} in the barrel region, outperforming the ECAL energy resolution in the range $\ET > 20\GeV$. Hence, it is assumed that the resonance shape is determined by the \ufxf cluster. Furthermore, the impact on the Gaussian core of the resonance shape from other contributions, such as $\PBpm\to\PJGy\Pgppm\to\EE\Pgppm$ with the wrong mass hypothesis on the track, and the partial reconstruction of $\PBz\to\PJGy\PKst\to\EE\PKp\PGpm$, is minor when the \EE invariant mass is near the \PJGy mass~\cite{BPH-22-005}.

Candidate $\PBpm\to\PJGy\PKpm\to\EE\PKpm$ decays are selected using the criteria described in Table~\ref{tab:BuCRselection}, in addition to the \PJGy candidate selection in Table~\ref{tab:JpsiCRselection}. A high-purity track with $\pt > 3.5\GeV$ and $\abs{\eta} < 2.4$ is selected. Also, the track needs to satisfy $\Delta z$ with the tag muon $<$0.1\cm, but $\DR > 0.1$ to disambiguate kaon candidates from the muon. The kaon candidate and electron tracks must form a vertex using the KF algorithm with a probability $>$1\%. The kaon candidate's 3D DCA to the vertex must be $<$0.1\cm. The transverse angle between the \PB meson candidate's momentum and the vector connecting the beam axis and the vertex (\alphaTwoD) must satisfy $\cos(\alphaTwoD) > 0.95$. The \DR between the \PJGy and the kaon candidate needs to be $<$0.8. The invariant masses of the kaon candidate and each electron are required to be $>$2\GeV to suppress contributions from \PD meson decays. Finally, the invariant mass of the \ufxf cluster and the kaon candidate must fulfill $4.5 < \mass(\PJGy,\PK) < 6\GeV$. If multiple \PB candidates satisfy all criteria for a given \PeME, the \PB candidate with the best vertex fit probability is selected.

\begin{table}[htb]
\centering
\topcaption{Selection criteria (in addition to Table~\ref{tab:JpsiCRselection}) for the $\PBpm\to\PJGy\PKpm\to\EE\PKpm$ control region.\label{tab:BuCRselection}}
\renewcommand{\arraystretch}{1.3}
\begin{scotch}{ll}
\multirow{6}{*}{kaon} & $\pt > 3.5\GeV$\\
& $\abs{\eta} < 2.4$\\
\cline{2-2}
& High-purity track\\
& Veto PF electrons and muons\\
\cline{2-2}
& $\Delta z(\PK,\PGm_{\mathrm{trig}}) < 0.1\cm$\\
& $\DR(\PK,\PGm_{\mathrm{trig}}) > 0.1$\\
\hline
\multirow{7}{*}{B meson} & Vertex fit probability $>$0.01\\
& $\textrm{DCA}(\PK,\textrm{vertex}) < 0.1\cm$\\
& $\cos{(\alphaTwoD)} > 0.95$\\
& $\DR(\PJGy, \PK) < 0.8$\\
\cline{2-2}
& $\mass(\Pe_{1},\PK) > 2\GeV$\\
& $\mass(\Pe_{2},\PK) > 2\GeV$\\
& $4.5 < \mass(\PJGy,\PK) < 6.0\GeV$\\
\end{scotch}
\end{table}

\begin{figure}[hbtp]
  \centering
    \includegraphics[width=0.49\textwidth]{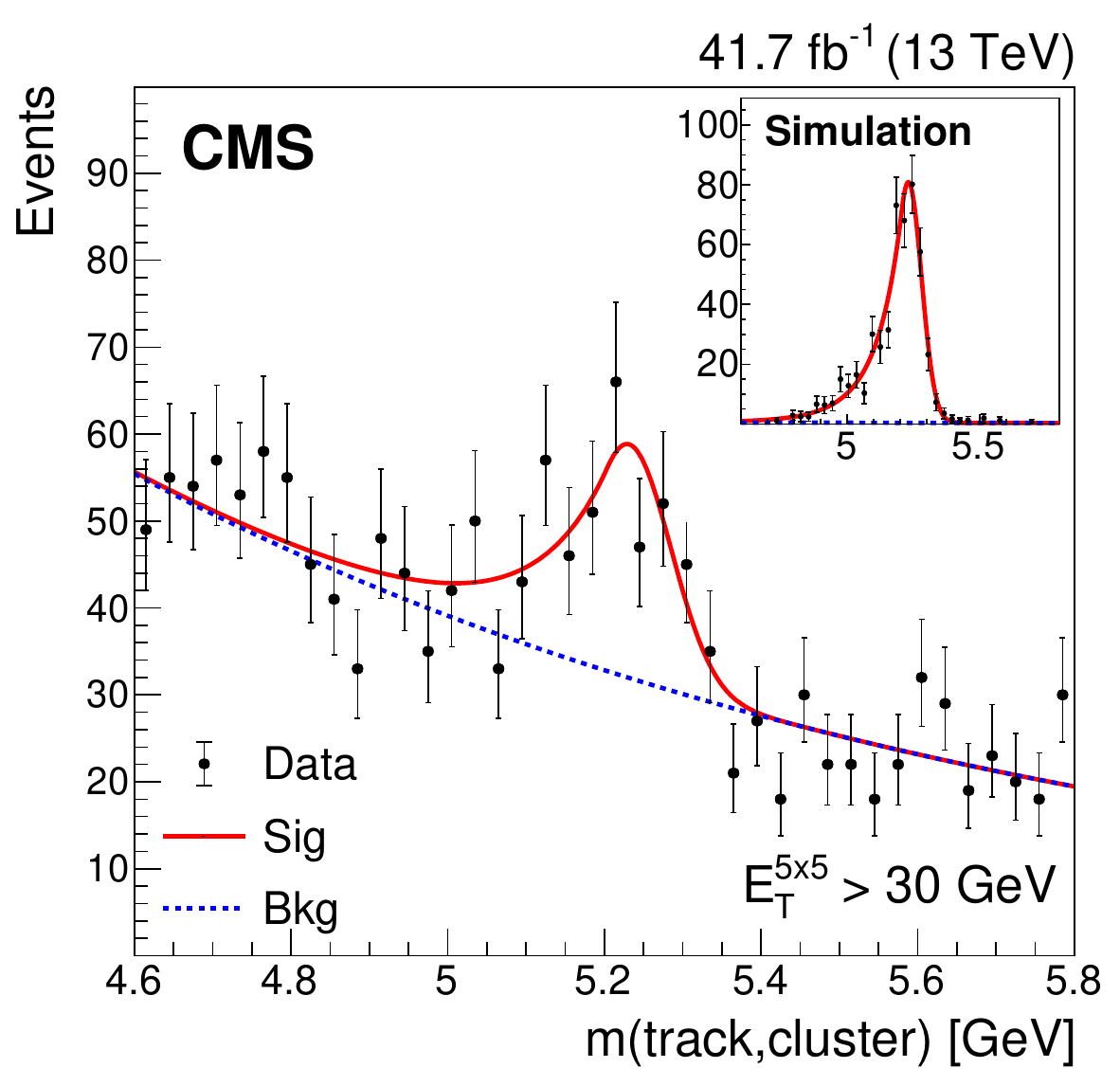}
    \caption{The distribution of the invariant mass between the \ufxf cluster and the kaon candidate with $\ETfxf > 30\GeV$. The vertical bar shows the statistical uncertainty. The signal (background) contribution is modeled with a Crystal Ball (exponential) function, represented with a red (blue) line. The inset on the top right illustrates the mass distribution in a $\PBpm\to\PJGy\PKpm\to\EE\PKpm$ MC sample.}
    \label{fig:mergedEle5x5CR}
\end{figure}

Figure~\ref{fig:mergedEle5x5CR} shows the invariant mass of the \ufxf cluster and the kaon candidate from data and simulation. The scale and resolution of the \PB meson resonance are extracted from the signal Crystal Ball function fit's mean and the standard deviation of its Gaussian core. The mean value is $5.231\pm0.006$ ($5.232\pm0.012$)~\GeV, while the standard deviation is $0.050\pm0.005$ ($0.053\pm0.010$)~\GeV in data (simulation). An energy correction to \ETfxf is introduced to equalize the energy scale and resolution in data and simulation. The energy correction parameters are defined in Eq.~(\ref{eq:enCorr}), 
\begin{equation}
{\ETfxf}_{\mathrm{corr}} = \alpha \ETfxf\left[ 1+\mathrm{Gaus}(0, \beta)\right],\label{eq:enCorr}
\end{equation} 
where $\mathrm{Gaus}(\mu,\sigma)$ is a Gaussian random number with the mean of $\mu$ and the standard deviation of $\sigma$. The energy scale correction parameter ($\alpha$) is estimated by matching the mean of the Crystal Ball function via scaling \ETfxf in data while fixing the MC shape and other fitted parameters. Likewise, the energy smearing correction parameter ($\beta$) is evaluated by equalizing the standard deviation of the Crystal Ball function by smearing \ETfxf in simulation, while fixing the data shape and other fitted parameters. The optimal values of $\alpha$ and $\beta$ to equalize the scale and resolution in data and simulation are found to be $1.003\pm0.003$ and $0.02\pm0.01$, respectively. As there are insufficient \PB candidates to evaluate the correction as a function of \ETfxf, correction parameters are derived inclusively. Therefore, the difference in the \ufxf cluster's energy before and after applying the scaling and smearing correction is assigned as a systematic uncertainty, assuming that the correction parameters are expected to be smaller in higher energy because the \ufxf energy resolution scales proportional to the inverse square root of \Eufxf.

\section{Summary}

Theories of beyond the standard model physics predict light bosons that subsequently decay into a dielectron (\EE). In such models, the light boson can be significantly boosted, and the standard clustering and reconstruction algorithm may fail to resolve the dielectron. The CMS standard electron reconstruction resolves about 5--40\% of \EE pairs when separated by \DR of 0.02--0.04, corresponding to 1--2 Moli{\`e}re radii, where \DR is defined as $\DR\equiv\sqrt{\smash[b]{(\Delta\eta)^{2}+(\Delta\phi)^{2}}}$. Moreover, high Lorentz boosts can lead to the absence of either of the electron tracks due to shared hits in the inner tracker. Therefore, novel algorithms to identify merged electrons (\PeME) are developed for the single-track and two-track cases, respectively.

The two-track model is trained using a boosted decision tree based on the geometrical compatibility between the merged cluster and the electron tracks. The default clustering algorithm in the electromagnetic calorimeter does not capture all energy deposits from \PeME candidates. A cluster consisting of a union of the $5\times5$ matrices of lead tungstate crystals around the two electron tracks (\ufxf) is used to estimate the energy of the \PeME instead of the default cluster for the two-track case. The \ufxf cluster's energy scale and resolution in data are measured with $\PBpm\to\PJGy\PKpm\to\EE\PKpm$ decays by reconstructing the invariant mass of the \ufxf cluster and a track, and found to be consistent with that of the simulation. The two-track model shows about 75\% efficiency for the \EE pairs with two reconstructed tracks separated by $\DR\simeq0.01$. The efficiency in data is validated using boosted \PJGy mesons decaying into \EE.

The single-track model targets \EE pairs with an extreme Lorentz boost ($\gammaL > 200$), and is mainly based on the ratio between the energy of the merged cluster and the reconstructed electron track. Its efficiency is about 50\% for the \EE pairs with a single reconstructed track at $\DR\simeq0.001$. The $\PZ\to\MM\PGg$ events, where the final-state radiation photon becomes the \EE pair with a single reconstructed track, are used to assess the efficiency in data, which is found to be 60\%.

\begin{acknowledgments}
We congratulate our colleagues in the CERN accelerator departments for the excellent performance of the LHC and thank the technical and administrative staffs at CERN and at other CMS institutes for their contributions to the success of the CMS effort. In addition, we gratefully acknowledge the computing centers and personnel of the Worldwide LHC Computing Grid and other centers for delivering so effectively the computing infrastructure essential to our analyses. Finally, we acknowledge the enduring support for the construction and operation of the LHC, the CMS detector, and the supporting computing infrastructure provided by the following funding agencies: SC (Armenia), BMBWF and FWF (Austria); FNRS and FWO (Belgium); CNPq, CAPES, FAPERJ, FAPERGS, and FAPESP (Brazil); MES and BNSF (Bulgaria); CERN; CAS, MoST, and NSFC (China); MINCIENCIAS (Colombia); MSES and CSF (Croatia); RIF (Cyprus); SENESCYT (Ecuador); ERC PRG and PSG, TARISTU24-TK10 and MoER TK202 (Estonia); Academy of Finland, MEC, and HIP (Finland); CEA and CNRS/IN2P3 (France); SRNSF (Georgia); BMFTR, DFG, and HGF (Germany); GSRI (Greece); MATE and NKFIH (Hungary); DAE and DST (India); IPM (Iran); SFI (Ireland); INFN (Italy); MSIT and NRF (Republic of Korea); MES (Latvia); LMTLT (Lithuania); MOE and UM (Malaysia); BUAP, CINVESTAV, CONACYT, LNS, SEP, and UASLP-FAI (Mexico); MOS (Montenegro); MBIE (New Zealand); PAEC (Pakistan); MSHE, NSC, and NAWA (Poland); FCT (Portugal); MESTD (Serbia); MICIU/AEI and PCTI (Spain); MOSTR (Sri Lanka); Swiss Funding Agencies (Switzerland); MST (Taipei); MHESI (Thailand); TUBITAK and TENMAK (T\"{u}rkiye); NASU (Ukraine); STFC (United Kingdom); DOE and NSF (USA).

\hyphenation{Rachada-pisek} Individuals have received support from the Marie-Curie program and the European Research Council and Horizon 2020 Grant, contract Nos.\ 675440, 724704, 752730, 758316, 765710, 824093, 101115353, 101002207, 101001205, and COST Action CA16108 (European Union); the Leventis Foundation; the Alfred P.\ Sloan Foundation; the Alexander von Humboldt Foundation; the Science Committee, project no. 22rl-037 (Armenia); the Fonds pour la Formation \`a la Recherche dans l'Industrie et dans l'Agriculture (FRIA) and Fonds voor Wetenschappelijk Onderzoek contract No. 1228724N (Belgium); the Beijing Municipal Science \& Technology Commission, No. Z191100007219010, the Fundamental Research Funds for the Central Universities, the Ministry of Science and Technology of China under Grant No. 2023YFA1605804, the Natural Science Foundation of China under Grant No. 12535004, and USTC Research Funds of the Double First-Class Initiative No.\ YD2030002017 (China); the Ministry of Education, Youth and Sports (MEYS) of the Czech Republic; the Shota Rustaveli National Science Foundation, grant FR-22-985 (Georgia); the Deutsche Forschungsgemeinschaft (DFG), among others, under Germany's Excellence Strategy -- EXC 2121 ``Quantum Universe" -- 390833306, and under project number 400140256 - GRK2497; the Hellenic Foundation for Research and Innovation (HFRI), Project Number 2288 (Greece); the Hungarian Academy of Sciences, the New National Excellence Program - \'UNKP, the NKFIH research grants K 131991, K 138136, K 143460, K 143477, K 147557, K 146913, K 146914, K 147048, TKP2021-NKTA-64, and 2025-1.1.5-NEMZ\_KI-2025-00004, and MATE KKP and KKPCs Research Excellence and Flagship Research Groups grants (Hungary); the Council of Science and Industrial Research, India; ICSC -- National Research Center for High Performance Computing, Big Data and Quantum Computing, FAIR -- Future Artificial Intelligence Research, and CUP I53D23001070006 (Mission 4 Component 1), funded by the NextGenerationEU program, the Italian Ministry of University and Research (MUR) under Bando PRIN 2022 -- CUP I53C24002390006, PRIN PRIMULA 2022RBYK7T (Italy); the Latvian Council of Science; the Ministry of Science and Higher Education, project no. 2022/WK/14, and the National Science Center, contracts Opus 2021/41/B/ST2/01369, 2021/43/B/ST2/01552, 2023/49/B/ST2/03273, and the NAWA contract BPN/PPO/2021/1/00011 (Poland); the Funda\c{c}\~ao para a Ci\^encia e a Tecnologia (Portugal); the National Priorities Research Program by Qatar National Research Fund; MICIU/AEI/10.13039/501100011033, ERDF/EU, ``European Union NextGenerationEU/PRTR", projects PID2022-142604OB-C21, PID2022-139519OB-C21, PID2023-147706NB-I00, PID2023-148896NB-I00, PID2023-146983NB-I00, PID2023-147115NB-I00, PID2023-148418NB-C41, PID2023-148418NB-C42, PID2023-148418NB-C43, PID2023-148418NB-C44, PID2024-158190NB-C22, RYC2021-033305-I, RYC2024-048719-I, CNS2023-144781, CNS2024-154769 and Plan de Ciencia, Tecnolog{\'i}a e Innovaci{\'o}n de Asturias, Spain; the Chulalongkorn Academic into Its 2nd Century Project Advancement Project, the National Science, Research and Innovation Fund program IND\_FF\_68\_369\_2300\_097, and the Program Management Unit for Human Resources \& Institutional Development, Research and Innovation, grant B39G680009 (Thailand); the Eric \& Wendy Schmidt Fund for Strategic Innovation through the CERN Next Generation Triggers project under grant agreement number SIF-2023-004; the Kavli Foundation; the Nvidia Corporation; the SuperMicro Corporation; the Welch Foundation, contract C-1845; and the Weston Havens Foundation (USA).
\end{acknowledgments}

\bibliography{auto_generated}   
\cleardoublepage \appendix\section{The CMS Collaboration \label{app:collab}}\begin{sloppypar}\hyphenpenalty=5000\widowpenalty=500\clubpenalty=5000\input{EGM-24-002-public-authorlist.tex}\end{sloppypar}
\end{document}

%% file: EGM-24-002-public-authorlist.tex
\cmsinstitute{Yerevan Physics Institute, Yerevan, Armenia}
{\tolerance=6000
A.~Hayrapetyan, V.~Makarenko\cmsorcid{0000-0002-8406-8605}, A.~Tumasyan\cmsAuthorMark{1}\cmsorcid{0009-0000-0684-6742}
\par}
\cmsinstitute{Institut f\"{u}r Hochenergiephysik, Vienna, Austria}
{\tolerance=6000
W.~Adam\cmsorcid{0000-0001-9099-4341}, J.W.~Andrejkovic, L.~Benato\cmsorcid{0000-0001-5135-7489}, T.~Bergauer\cmsorcid{0000-0002-5786-0293}, M.~Dragicevic\cmsorcid{0000-0003-1967-6783}, C.~Giordano\cmsorcid{0000-0001-6317-2481}, P.S.~Hussain\cmsorcid{0000-0002-4825-5278}, M.~Jeitler\cmsAuthorMark{2}\cmsorcid{0000-0002-5141-9560}, N.~Krammer\cmsorcid{0000-0002-0548-0985}, A.~Li\cmsorcid{0000-0002-4547-116X}, D.~Liko\cmsorcid{0000-0002-3380-473X}, M.~Matthewman, I.~Mikulec\cmsorcid{0000-0003-0385-2746}, J.~Schieck\cmsAuthorMark{2}\cmsorcid{0000-0002-1058-8093}, R.~Sch\"{o}fbeck\cmsAuthorMark{2}\cmsorcid{0000-0002-2332-8784}, D.~Schwarz\cmsorcid{0000-0002-3821-7331}, M.~Shooshtari\cmsorcid{0009-0004-8882-4887}, M.~Sonawane\cmsorcid{0000-0003-0510-7010}, W.~Waltenberger\cmsorcid{0000-0002-6215-7228}, C.-E.~Wulz\cmsAuthorMark{2}\cmsorcid{0000-0001-9226-5812}
% ...
\par}
\vskip\cmsinstskip
\dag:~Deceased\\
$^{1}$Also at Yerevan State University, Yerevan, Armenia\\
$^{2}$Also at TU Wien, Vienna, Austria\\
% ...
